\documentclass[12pt]{article}
\usepackage{amsmath}
\usepackage{latexsym}
\usepackage{bbm}
\usepackage{amssymb}

\def\a{\alpha}

\def\b{\beta}
\def\g{\gamma}

\def\d{\delta}

\def\e{\varepsilon}

\def\z{\zeta}

\def\th{\theta}

\def\k{\kappa}
\def\l{\lambda}
\def\m{\mu}
\def\n{\nu}
\def\o{\omega}

\def\r{\rho}

\def\s{\sigma}

\def\td{\theta^{\dagger}}

\def\ed{\e^{\dagger}}

\def\rd{\r^{\dagger}}

\def\Yd{Y^{-1}}
\def\Zd{Z^{-1}}

\def\G{\Gamma}
\def\D{\Delta}

\def\S{\Sigma}

\def\L{\Lambda}
\def\O{\Omega}

\def\pa{\partial}

\def\half{\frac{1}{2}}

\def\tr{{\rm tr}}

\def\Str{{\rm str}}
\def\and{{\rm and}}

\def\ie{{\it i.e.,} }

\def\IR{\mathbbm R}
\def\IZ{\mathbbm Z}

\def\cA{{\cal A}}

\begin{document}
\vspace*{-1.0in}
\thispagestyle{empty}
\begin{flushright}
CALT-TH-2015-030
\end{flushright}

%\large
\normalsize
\baselineskip = 18pt
\parskip = 6pt

\vspace{1.0in}

{\Large \begin{center}
{\bf New Formulation of the Type IIB\\ Superstring Action in $\mathbf{AdS_5 \times S^5}$}
\end{center}}

\vspace{.25in}

\begin{center}
John H. Schwarz\footnote{jhs@theory.caltech.edu}
\\
\emph{Walter Burke Institute for Theoretical Physics\\
California Institute of Technology 452-48\\ Pasadena, CA  91125, USA}
\end{center}
\vspace{.25in}

\begin{center}
\textbf{Abstract}
\end{center}
\begin{quotation}

Previous studies of the type IIB superstring in an ${AdS_5 \times S^5}$
background are based on a description of the superspace geometry as the quotient
space $PSU(2,2|4)/SO(4,1) \times SO(5)$.
This paper develops an alternative approach in which the Grassmann coordinates
provide a nonlinear realization of $PSU(2,2|4)$ based on the quotient space
$PSU(2,2|4)/SU(2,2) \times SU(4)$, and the bosonic coordinates are described
as a submanifold of $SU(2,2) \times SU(4)$. This formulation
is used to construct the superstring world-sheet action in a form in which the
$PSU(2,2|4)$ symmetry is manifest and local kappa symmetry can be established.
It provides the complete dependence on the Grassmann coordinates in terms of
simple analytic expressions. Therefore it is expected to have advantages compared
previous approaches, but this remains to be demonstrated.

\end{quotation}

\newpage

\pagenumbering{arabic}

\tableofcontents

\newpage

\section{Introduction}

The conjectured duality \cite{Maldacena:1997re} between type IIB superstring theory \cite{Green:1981yb}
in a maximally supersymmetric ${AdS_5 \times S^5}$ background,
with $N$ units of self-dual five-form flux,
and four-dimensional ${\cal N} =4$ super Yang--Mills theory \cite{Brink:1976bc},
with a $U(N)$ gauge group,
has been studied extensively. This is a precisely defined conjecture, because
the ${AdS_5 \times S^5}$ background is an exact solution
of type IIB superstring theory \cite{Kallosh:1998qs}.
Most studies have focused on the large-$N$
limit for fixed 't Hooft parameter $\l = g_{YM}^2 N$.
(See \cite{Beisert:2010jr} and references therein.)
This limit corresponds to the planar approximation to the field theory \cite{'tHooft:1973jz}
and the classical (or leading genus) approximation to the string theory. The
planar approximation to the field theory is an integrable four-dimensional theory,
with an infinite-dimensional Yangian symmetry generated by the superconformal group
$PSU(2,2|4)$ and a dual conformal group. Its perturbative expansion parameter
is proportional to $\l$.

The isometry supergroup of the ${AdS_5 \times S^5}$ solution of type IIB superstring theory is
also $PSU(2,2|4)$. Its bosonic subgroup is $SU(4) \times SU(2,2)$, where $SU(4) = Spin(6)$
and $SU(2,2) = Spin(4,2)$. This supergroup has 32 fermionic generators,
which we will refer to as supersymmetries. This is the maximum number possible and
the same number as the flat-space solution, which corresponds to the large-radius
(or large-$\l$) limit of the $AdS_5 \times S^5$ solution.
The string theory, for the background in question, is described by an interacting two-dimensional
world-sheet theory, whose perturbative expansion parameter is proportional to $1/\sqrt{\l}$. This
theory is also integrable.

Even though the planar ${\cal N}=4$ super Yang--Mills theory and the leading-genus
$AdS_5 \times S^5$ superstring action are both integrable, both of them are also very challenging
to study. Nonetheless, a lot of progress has been achieved, providing convincing evidence
in support of the duality, thanks to an enormous effort by many very clever people.
The goal of the present work is to derive a new formulation of the superstring
world-sheet theory. It will turn out to be equivalent to
the previous formulation by Metsaev and Tseytlin \cite{Metsaev:1998it}
and others \cite{Kallosh:1998zx} \cite{Pesando:1998fv}.\footnote{For a
recent review, see \cite{Arutyunov:2009ga}.}
However, it has some attractive features that might make it more useful.
%enable more tractable calculations,
%at least in some cases, and possibly suggest new insights.

In recent work the author has studied the bosonic truncation of the
world-volume action of a probe D3-brane embedded in this background and made
certain conjectures concerning an interpretation of this action
that should hold when the fermionic
degrees of freedom are incorporated \cite{Schwarz:2013wra}\cite{Schwarz:2014rxa}.
This provided the motivation for developing a convenient formalism for
adding the fermions in which all of the symmetries can be easily understood.
While that is the motivation, the present work does not require the
reader to be familiar with those papers, nor does it depend on the correctness of
their conjectures, which have aroused considerable skepticism.
This paper, which is about superspace geometry and the superstring action, does not
make any bold conjectures, and therefore it should be noncontroversial.

String world-sheet actions have much in common with WZW models for groups, supergroups,
cosets, etc. though there a few differences. One difference is that they are invariant
under reparametrization of the world-sheet coordinates. One way of implementing this is
to couple the sigma model to two-dimensional gravity. If one chooses a conformally
flat gauge, the action reduces to the usual two-dimensional
Minkowski space form, but it is supplemented by
Virasoro constraints. Residual symmetries in this gauge allow further gauge fixing,
the main example being light-cone gauge.
In addition to the local reparametrization invariance, superstring actions also have
local fermionic symmetries, called kappa symmetry. They are rather subtle, and
they play a crucial role. One of the goals of this paper is to give a clear
explanation of how kappa symmetry is realized.

In constructing chiral sigma models for homogeneous spaces that have an isometry group $G$,
but are not group manifolds, the standard approach is to formulate them as coset
theories. Thus, for example, a theory on a sphere $S^n$ is formulated as an $SO(n+1)/SO(n)$
coset theory. The formulas that describe symmetric spaces as
$M =G/H$ coset theories are well-known \cite{Coleman:1969sm}\cite{Callan:1969sn}.
They involve a construction that incorporates
global $G$ symmetry and local $H$ symmetry. This is the standard thing
to do, and so it is not surprising that this is the approach that was utilized in
\cite{Metsaev:1998it} to construct the superstring world-sheet action for
$AdS_5 \times S^5$. In this case the coset in question is
$PSU(2,2|4)/SO(4,1) \times SO(5)$.
This paper describes an alternative procedure in which the Grassmann coordinates
provide a nonlinear realization of $PSU(2,2|4)$ based on the quotient space
$PSU(2,2|4)/SU(2,2) \times SU(4)$, and the bosonic coordinates are described
as a submanifold of $SU(2,2) \times SU(4)$.

The description of $S^2$ as a subspace of $SU(2)$ is a very simple analog of the
procedure that will be used. (It is relevant to the discussion of $AdS_2 \times S^2$,
which is analogous to $AdS_5 \times S^5$.) The group
$SU(2)$ consists of $2 \times 2$ unitary unimodular matrices, and the group
manifold is $S^3$. An $S^2$ can be embedded in this group manifold in many different
ways. The one that is most relevant for our purposes is the subspace consisting
of all symmetric $SU(2)$ matrices. This subspace can be expressed in
terms of Pauli matrices in the form $\s_2 \vec\s\cdot\hat x$, where $\hat x$ is
a unit-length 3-vector. The action of a group element $g$ on
an element of this sphere, represented by a symmetric matrix $g_0$, is
$ g_0 \to g^T g_0 g$. This is a point on the same $S^2$, since $g^T g_0 g$ is also
a symmetric $SU(2)$ matrix. The isometry group of $S^2$ is actually $SO(3)$,
because the group elements $g$ and $-g$ describe the same map.
Clearly, a specific subspace of $SU(2)$ has been selected, in a way
that does not depend on any arbitrary choices, to describe $S^2$.
This paper applies a similar procedure to the description of $S^5$ and $AdS_5$
as subspaces of $SU(4)$ and $SU(2,2)$, respectively. In particular, the
description of $S^5$ in terms of antisymmetric $SU(4)$ matrices is discussed
in detail in Appendix~A.

This formulation of the superspace geometry
that enters in the construction of the superstring action
makes it possible to keep all of the bosonic symmetries manifest throughout
the analysis,\footnote{A previous attempt to make
the $SU(4)$ symmetry manifest is described in \cite{Metsaev:2000ds}.} and
many formulas, including the superstring action itself, have manifest $PSU(2,2|4)$
symmetry. Also, the complete dependence on the Grassmann coordinates for
all relevant quantities is given by simple tractable analytic expressions.
So far, we have just rederived results that have been known for a long time,
but the hope is that this reformulation of the world-sheet theory
will be helpful for obtaining new results.

\section{The bosonic truncation}

Before confronting superspace geometry, let us briefly review
the bosonic structure of $AdS_5 \times S^5$, which has the isometry $SO(4,2) \times SO(6)$.
The generators of $SO(6)$, denoted $J^{ab}$, where $a,b = 1,2,\ldots6$,
can be viewed as generators of rotations of $\IR^{6}$ about the origin.
They also generate the isometries of a unit-radius $S^5$ centered about the origin
($\hat z \cdot \hat z = \sum_1^{6} (z^a)^2 = 1$).
Similarly, $J^{mn}$ generates isometries of a unit-radius $AdS_{5}$ embedded in
$\IR^{4,2}$ by the equation
\begin{equation} \label{AdS5}
\hat y \cdot \hat y = \sum_{m,n =0}^5 \eta_{mn} y^m y^n =
-(y^0)^2 + (y^1)^2 + (y^2)^2 + (y^3)^2 +(y^4)^2 - (y^5)^2 = -1.
\end{equation}
This equation describes the Poincar\'e patch of $AdS_5$, which is all that we are concerned
with in this work. The two algebras are distinguished by the choice of indices
($a,b,c,d$ or $m,n,p,q$).

We prefer to write the unit-radius ${AdS_5 \times S^5}$
metric in a form in which all of the isometries
are manifest. There are various ways to achieve this. One
option is
\begin{equation}\label{bmetric}
ds^2 = d\hat z \cdot d \hat z + d\hat y \cdot d \hat y,
\end{equation}
where $\hat z$ and $\hat y$ are understood to satisfy the constraints described above.
This is the description that will be utilized in most of this paper.

Lie-algebra-valued connection one-forms associated to the $SO(6)$ symmetry of $S^5$ are easily
constructed in terms of the unit six-vector $z^a$. (We do not display hats to avoid clutter.)
The one-form is
\begin{equation}
\O_0^{ab} = 2(z^a dz^b - z^b dz^a).
\end{equation}
The subscript 0 is used to refer to the bosonic truncation. The normalization is chosen
to ensure that this is a flat connection, \ie its two-form curvature is
\begin{equation}
d\O_0^{ab} + \O_0^{ac}\wedge  \eta_{cd}\, \O_0^{db} = 0.
\end{equation}
This is easily verified using $z^a\, \eta_{ab}\, dz^b = 0$, which is a consequence
of $z^2 = z^a \,\eta_{ab} \, z^b = 1$. In the case of $SO(6)$, the metric $\eta$ is just a
$6 \times 6$ unit matrix, which we denote $I_6$. Similarly, there is
a Lie-algebra-valued flat connection
\begin{equation}
\tilde\O_0^{mn} = -2(y^m dy^n - y^n dy^m)
\end{equation}
associated to the $SO(4,2)$ symmetry of $AdS_5$. In the $SO(4,2)$ case  $\eta = I_{4,2}$, which has diagonal components $(1,1,1,1,-1,-1)$. Recall that for this choice $y^2 = -1$, which is why
$\tilde\O_0^{mn}$ requires an extra minus sign to ensure flatness.

The bosonic truncation of the superstring action can be expressed
entirely in terms of the induced world-volume metric,
\begin{equation}
 G_{\a\b} = \pa_\a \hat z \cdot \pa_\b \hat z + \pa_\a \hat y \cdot \pa_\b \hat y,
\end{equation}
where it is understood that $y$ and $z$ are functions of the world-sheet
coordinates $\s^\a$, $\a=0,1$. The action is
\begin{equation}
S = - \frac{R^2}{2 \pi \a'} \int d^2 \s \sqrt{-G},
\end{equation}
where $G = \det G_{\a\b}$ and $\a'$ is the usual string theory Regge slope parameter,
which (for $\hbar = c = 1$) has dimensions of length squared. $R$ is the radius of both
$S^5$ and $AdS_5$. A standard rewriting of such
a metric involves introducing an auxiliary world-sheet metric field $h_{\a\b}$. Then
the action can be recast as
\begin{equation}
S = - \frac{R^2}{4 \pi \a'} \int d^2 \s \sqrt{-h} h^{\a\b} G_{\a\b}.
\end{equation}
This form has a Weyl symmetry given by an arbitrary local rescaling of $h_{\a\b}$.
The simplest way to understand the equivalence of the two forms of $S$ is to note
that the $h_{\a\b}$ classical equation of motion is solved by $h_{\a\b} = e^{f(\s)} G_{\a\b}$,
\ie they are conformally equivalent. The conformal factor cancels out classically.
For a critical string theory, without conformal anomaly, it should also cancel
quantum mechanically. The bosonic truncation described here is not critical,
but the complete theory should be.

When the fermionic degrees of freedom are included,
the dual CFT is ${\cal N} = 4$ super Yang--Mills theory with
a $U(N)$ gauge group. $N$ is related to a five-form flux, which does not
appear in the string world-sheet action. (It does appear in the D3-brane action.)
The gauge theory has a dimensionless
't Hooft parameter $\l =g_{YM}^2 N$. AdS/CFT duality gives the identification
\begin{equation}
g_s = \frac{g_{YM}^2}{4\pi},
\end{equation}
where $g_s$ is the string coupling constant (determined by the vev of the dilaton field).
The radius $R$ of the $S^5$ and the $AdS_5$ is introduced by replacing the unit-radius
metric $ds^2$ by $R^2 ds^2$. Then, utilizing the AdS/CFT identification
\begin{equation}
R^2 = \a'\sqrt{\l},
\end{equation}
one obtains
\begin{equation}
S = - \frac{\sqrt{\l}}{2\pi}\int d^2 \s \sqrt{-G} .
\end{equation}
In the large $N$ limit, taken at fixed $\l$, the CFT is described by the planar
approximation, and the string theory is described by the classical approximation, \ie
leading order in the world-sheet genus expansion, which is a cylinder. Even so,
the two-dimensional world-sheet theory must be treated as a quantum theory, with a
large-$\l$ perturbation expansion in powers of $1/\sqrt{\l}$, in order to determine
the string spectrum and tree amplitudes. Flat ten-dimensional spacetime is the
leading approximation in this expansion. The dual planar CFT, on the other hand,
has a small-$\l$ perturbation expansion in powers of $\l$.

Let us introduce null world-sheet coordinates $\s^{\pm} = \s^1 \pm \s^0$.
It is often convenient to choose a conformally flat gauge. This means using the
two diffeomorphism symmetries to set
\begin{equation}
G_{++} = G_{--} =0.
\end{equation}
Then the action simplifies to
\begin{equation}
S = - \frac{\sqrt{\l}}{2\pi}\int d^2 \s G_{+-},
\end{equation}
which is supplemented by the Virasoro constraints $G_{++} = G_{--} =0$.\footnote{When the
world-sheet theory is quantized, $G_{++}$ and $G_{--}$ become operators that need
to be treated with care. In any case, the bosonic truncation of the world-sheet theory
is inconsistent beyond the classical approximation due to a conformal anomaly.}
In the geometry at hand, we have
\begin{equation}
 G_{+-} = \pa_+ \hat z \cdot \pa_- \hat z + \pa_+ \hat y \cdot \pa_- \hat y.
\end{equation}
This can then be varied to give equations of motion. Taking account of the constraints
$\hat z \cdot \hat z = 1$ and $\hat y \cdot \hat y = -1$, we obtain
\begin{equation}
(\eta^{ab} - z^a z^b) \pa_+ \pa_- z_b = 0 \quad {\rm and} \quad
(\eta^{mn} + y^m y^n) \pa_+ \pa_- y_n = 0.
\end{equation}
Conservation of the $SO(6)$ and $SO(4,2)$ Noether currents implies that
\begin{equation}
\pa_\a(z^a \pa^\a z^b -  z^b \pa^\a z^a) = 0 \quad {\rm and} \quad
\pa_\a(y^m \pa^\a y^n -  y^n \pa^\a y^m) = 0 .
\end{equation}
These 30 equations are consequences of the 10 preceding equations. In fact, they are
equivalent to them. Expressed more elegantly, they take the form
\begin{equation}
d\star\O_0^{ab} = 0 \quad {\rm and} \quad
d \star \tilde\O_0^{mn} =0.
\end{equation}

The bosonic connections $\O_0$ and $\tilde\O_0$ are simultaneously conserved and flat
when the equations of motion are taken into account. These conditions allow one
to construct a one-parameter family of flat connections, whose existence is the key to classical
integrability of the world-sheet theory \cite{Mandal:2002fs}.
In the remainder of this manuscript we will add Grassmann
coordinates and construct the complete superstring action with $PSU(2,2|4)$ symmetry.
Since this will be a ``critical'' string theory (without conformal anomaly), its
integrability is expected to be valid for the quantum theory, \ie taking full account of
the dependence on $\l$, but only at leading order in the genus expansion.

\section{Supersymmetrization}

Our goal is to add fermionic (Grassmann) coordinates $\th$ to the metric of the preceding
section so as to make it invariant under ${PSU(2,2|4)}$.
In addition to bosonic one-forms $\O^{ab}$ and $\tilde\O^{mn}$, whose bosonic
truncations are $\O_0^{ab}$ and $\tilde\O_0^{mn}$ described in Sect.~2, we also require
a fermionic one-form $\Psi$, which is dual to the fermionic supersymmetry generators of the
superalgebra. $\Psi$ and $\Psi^{\dagger}$ should encode 32 fermionic one-forms,
which transform under $SU(4)\times SU(2,2)$ as ${\bf (4, \bar4)} + {\bf (\bar4,4)}$.
%\footnote{Whether
%one calls them ${\bf (4, \bar4)} + {\bf (\bar4,4)}$ or ${\bf (4, 4)} + {\bf (\bar4, \bar4)}$
%is a matter of convention.}

Let us recast the connections $\O$ and $\tilde\O$ in spinor notation.
The construction for $SU(4)$ requires $4 \times 4$ analogs of Pauli matrices, or Dirac matrices,
denoted $\S^a$, which are described in Appendix~A. In the notation described there,
we define
\begin{equation}
\O^{\a}{}_{\b}
%=  (\S_{ab})^{\a}{}_{\b} \hat z^a d\hat z^b
= \frac{1}{4} (\S_{ab})^{\a}{}_{\b} \O^{ab}.
\end{equation}
Also in the notation described in Appendix~A, there is an identical-looking
formula for $SU(2,2)$,
\begin{equation}
\tilde\O^{\m}{}_{\n}
%=  (\S_{mn})^{\m}{}_{\n} \hat y^m d\hat y^n
= \frac{1}{4} (\S_{mn})^{\m}{}_{\n} \tilde\O^{mn}.
\end{equation}
Infinitesimal parameters of $SU(4)$ and $SU(2,2)$ transformations are described
in spinor notation by matrices $\o^{\a}{}_{\b}$ and $\tilde\o^{\m}{}_{\n}$ in an analogous manner.

The fermionic one-form, transforming as ${\bf (4, \bar4)}$, is written $\Psi^{\a}{}_{\m}$.
Its hermitian conjugate, which transforms as ${\bf (\bar4,4)}$
is written $(\Psi^\dagger)^{\m}{}_{\a}$.
Spinor indices can be lowered or contracted using the $4 \times4$ invariant
tensors $\eta_{\a\bar \b}$ and $\eta_{\m \bar\n}$. $\eta_{\a\bar \b}$ is just the unit matrix $I_4$,
and $\eta_{\m \bar\n}$ is the $SU(2,2)$ metric $I_{2,2}$. Thus, for example,
$\Psi^{\a \bar\m}= \Psi^{\a}{}_{\n} \eta^{\n \bar \m}$.
By always using matrices with unbarred indices we avoid the need to ever display $\eta$
matrices explicitly. The price one pays for this is that expressions that are called adjoints,
such as $\Psi^{\dagger}$, are not conventional adjoints, since they contain additional $\eta$
factors. However, this ``adjoint'' is still an involution, since the square of an $\eta$ is
a unit matrix. In this notation, it makes sense to call the matrix $\tilde\O$ antihermitian
despite the indefinite signature of $SU(2,2)$.

\subsection{Supermatrices}

Since it is convenient to represent supergroups using supermatrices, let us review a
few basic facts and our conventions. There
are various conventions in the literature, and we shall introduce yet
another one. We write an $8 \times 8$ supermatrix in terms of
$4 \times 4$ blocks as follows
\begin{equation}
M
= \left( \begin{array}{cc}
a &  \z b  \\
\z c & d \\
\end{array} \right) ,
\end{equation}
where $a$ and $d$ are Grassmann even and $b$ and $c$ are Grassmann odd.
$a$ is the $SU(4)$ block and $d$ is the $SU(2,2)$ block.
This formula contains the phase
\begin{equation}
\z =  e^{-i\pi/4} ,
\end{equation}
which satisfies $\z^2 = -i$. By introducing factors of $\z$ in this way
various formulas have a more symmetrical appearance than is the case for
other conventions.

The {\em superadjoint} is defined by
\begin{equation} \label{superadjoint}
M^{\dagger}
= \left( \begin{array}{cc}
a^{\dagger} &  -\z c^{\dagger}  \\
-\z b^{\dagger} & d^{\dagger} \\
\end{array} \right) .
\end{equation}
This definition, which reduces to the usual one for the even blocks,
is chosen to ensure the identity
$(M_1 M_2)^{\dagger} = M_2^{\dagger} M_1^{\dagger}$.
By definition, a unitary supermatrix satisfies $M M^{\dagger} = I$ and an antihermitian
supermatrix satisfies $M + M^{\dagger} =0$.
Similarly, the {\em supertranspose} is defined by
\begin{equation} \label{supertranspose}
M^T
= \left( \begin{array}{cc}
a^T &  - i\z c^T  \\
-i\z b^T & d^T \\
\end{array} \right) .
\end{equation}
This satisfies $ (M_1 M_2)^T = M_2^T M_1^T$. However, it has the somewhat surprising property
\begin{equation}
(M^T)^T =  \left( \begin{array}{cc}
a &  -\z b  \\
-\z c & d \\
\end{array} \right) ,
\end{equation}
which makes the supertranspose a $\IZ_4$ transformation \cite{Berkovits:1999zq}.
Note that in verifying these formulas,
it is important to use the rules $ (bc)^{\dagger} = - c^{\dagger} b^{\dagger}$ and
$ (bc)^T = - c^T b^T$ for Grassmann odd matrices. It will also be useful to
define the inverse transpose, which has the same properties as the transpose,
\begin{equation} \label{barT}
M^{\overline T}
= \left( \begin{array}{cc}
a^T &   i\z c^T  \\
i\z b^T & d^T \\
\end{array} \right) .
\end{equation}
Which is which is a matter of convention.

The {\em supertrace} is defined by
\begin{equation}
\Str M = \tr\, a - \tr \, d .
\end{equation}
One virtue of this definition is that the familiar identity
$\tr (a_1 a_2) = \tr (a_2 a_1)$ generalizes to
\begin{equation}
\Str (M_1 M_2) = \Str (M_2 M_1).
\end{equation}
Another virtue is that $\Str M^T = \Str M^{\overline T} = \Str M$.

Our main concern in this work is the supergroup $PSU(2,2|4)$. The corresponding
superalgebra is best described in terms of matrices belonging to the superalgebra
$\mathfrak{su}(2,2|4)$. This algebra consists of antihermitian supermatrices with vanishing
supertrace. (It is implicit here that one takes appropriate account of the indefinite
signature of $\mathfrak{su}(2,2)$.) Given this algebra, one defines the
$\mathfrak{psu}(2,2|4)$ algebra
to consist of $\mathfrak{su}(2,2|4)$ matrices modded out by the equivalence relation
$M \sim M + i\l I$, where $I$ denotes the unit supermatrix. The factor of $i$ is shown because
$\l$ is assumed to be real and $M$ is supposed to be antihermitian.

\subsection{Nonlinear realization of the superalgebra}

Superspace is described by the bosonic spacetime coordinates $y^m$ and $z^a$,
satisfying $z^2 = 1$ and $y^2 = -1$, introduced in Sect.~2,
and Grassmann coordinates $\th^\a{}_\m$.
The $\th$ coordinates are 16 complex Grassmann numbers
that transform under $SU(4)\times SU(2,2)$ as ${\bf (4, \bar4)}$, like the one-form $\Psi$
discussed above. It will be extremely helpful to
think of $\th$ as a $4 \times 4$ matrix rather than as a 16-component spinor.
The two points of view are equivalent, of course, but matrix notation
will lead to much more elegant formulas. If all matrix multiplications were done
from one side, a tensor product notation (like that in Appendix~C) would be required.
Using matrix notation, we will obtain simple analytic
expressions describing the full $\th$ dependence of all quantities that
are required to formulate the superstring action.

One clue to understanding the $PSU(2,2|4)$ symmetry of the $AdS_5 \times S^5$
geometry is its relationship to the super-Poincar\'e symmetry
algebra of flat ten-dimensional superspace, which corresponds to the large-radius limit.
The large-radius limit preserves all 32 fermionic symmetries, but it only accounts for
30 of the 55 bosonic symmetries of the Poincar\'e algebra in 10 dimensions. The 25
rotations and Lorentz transformations that relate the $\IR^{4,1}$ piece of the
geometry that descends from $AdS_5$ to the $\IR^5$ piece that descends from $S^5$
are additional ``accidental'' symmetries of the limit.

A little emphasized feature
of the superspace description of the flat-space geometry is that the entire
super-Poincar\'e algebra closes on the Grassmann coordinates $\th$. A possible
reason for this lack of emphasis may be that the ten spacetime translations
act trivially, \ie they leave $\th$ invariant.
We will demonstrate here that the entire $\mathfrak{psu}(2,2|4)$ superalgebra closes on
the fermionic coordinates $\th^\a{}_\m$ even though the radius is finite.
In this case all of the symmetries transform $\th$ nontrivially, and
none of the transformations of $\th$ give
rise to expressions involving the $y$ or $z$ coordinates. This means that the
Grassmann coordinates provide a nonlinear realization of the superalgebra. Conceptually,
this is similar to the way the supersymmetry of a field theory in flat spacetime
is realized nonlinearly on a spinor field (the Goldstino) \cite{Volkov:1973ix}. In fact, the
algebra for the two problems is quite similar. The nonlinear Lagrangian for the
Goldstino field was generalized to anti de Sitter space in \cite{Zumino:1977av}.
However, that work is not directly relevant, since the goal of the present work
is to describe world-sheet fields and not ten-dimensional target-space fields.
The latter may deserve further consideration in the future.

The infinitesimal bosonic symmetry transformations of $\th$ are relatively trivial;
they are ``manifest'' in the sense that they are determined by the types
of spinor indices that appear. In matrix notation,
\begin{equation}
\d\th^\a{}_\m = (\o\th - \th \tilde\o )^\a{}_\m.
\end{equation}
The infinitesimal parameters $\o^\a{}_\b$ and ${\tilde\o}^\m{}_\n$  take values in
the $\mathfrak{su}(4)$ and $\mathfrak{su}(2,2)$ Lie algebras, respectively.
Thus, they are anti-hermitian
(in the sense discussed earlier) and traceless.

Let us now consider an infinitesimal supersymmetry
transformation of $\th$. In the case of flat space this is just $\d \th = \e$,
where $\e$ is an infinitesimal constant matrix of complex Grassmann parameters. In the
case of unit radius it is a bit more interesting:\footnote{This formula has appeared
previously in \cite{Berkovits:2007zk}\cite{Berkovits:2007rj}.}
\begin{equation}\label{deltath}
\d\th^\a{}_\m = \e^\a{}_\m + i(\th\ed \th)^\a{}_\m .
\end{equation}
The hermitian conjugate equation is then
\begin{equation}
\d(\td)^\m{}_\a = (\ed)^\m{}_\a + i(\td\e \td)^\m{}_\a .
\end{equation}
We have displayed the spinor indices, but the more compact formulas
$\d\th = \e + i\th\e^{\dagger} \th$ and
$\d\td = \e^{\dagger} + i\td\e \td$ are completely unambiguous. In our notation,
the quantities
\begin{equation} \label{udefs}
u = i \th \td \quad {\rm and} \quad \tilde u = i \td\th
\end{equation}
are both hermitian. 

In our conventions all coordinates ($\th, y, z$) are dimensionless, since they pertain to
unit radius ($R=1$). If we were to give them the usual dimensions, by absorbing appropriate
powers of $R$, then the second term in $\d\th = \e + i\th\e^{\dagger} \th$ 
would contain a coefficient $1/R$.
This makes it clear that this term vanishes in the large-radius limit.

It is a beautiful exercise to compute the commutator of two of these
supersymmetry transformations,
\[
[\d_1, \d_2]\th = \d_1(\e_2 +i \th \e_2^{\dagger}\th) - (1\leftrightarrow2)
\]
\begin{equation}\label{comm}
=i(\e_1 + i\th \e_1^{\dagger}\th)\e_2^{\dagger}\th
+ i\th \e_2^{\dagger}(\e_1 + i\th \e_1^{\dagger}\th)
- (1\leftrightarrow2)
\end{equation}
\[
=\o_{12} \th - \th {\tilde\o}_{12},
\]
where $\o_{12}$ and ${\tilde\o}_{12}$ are
\begin{equation} \label{omegaS}
(\o_{12})^\a{}_\b = i(\e_1\e_2^{\dagger} - \e_2\e_1^{\dagger})^\a{}_\b - {\rm trace},
%-\frac{1}{4} \tr(\e_2\e_1^{\dagger} + \e_2^{\dagger} \e_1) \d^\a_\b,
\end{equation}
\begin{equation} \label{omegaA}
(\tilde\o_{12})^\m{}_\n = i(\e_1^{\dagger}\e_2 - \e_2^{\dagger}\e_1)^\m{}_\n -{\rm trace}.
%-\frac{1}{4} \tr(\e_2\e_1^{\dagger} + \e_2^{\dagger} \e_1) \d^\m_\n.
\end{equation}
These are antihermitian, as required, since
$(\e_1 \e_2^{\dagger})^{\dagger} = - \e_2 \e_1^{\dagger}$.
Traces are subtracted in order that they are Lie-algebra valued. This is possible
due to the fact that the two trace terms give canceling contributions to Eq.~(\ref{comm}).
This commutator is exactly what the superalgebra requires it to be, demonstrating
that $\mathfrak{psu}(2,2|4)$ is nonlinearly realized entirely in terms of the Grassmann coordinates.

The transformation rule in Eq.~(\ref{deltath}) is not a unique
choice. The nonuniqueness corresponds to the possibility of redefining $\th$ by
introducing $\th' = \th + i c_1 \th\td\th  + \ldots$.
Then, the transformation rule would be modified accordingly. One could
even incorporate $y$ and $z$ in a redefinition, which would be truly perverse.
The choice that we have made is clearly the simplest and most natural one,
so it will be used in the remainder of this work.

It is possible to construct elements of the supergroup, represented by unitary supermatrices,
which are constructed entirely out of the Grassmann coordinates. For this purpose,
let us consider the supermatrix\footnote{This description was suggested by W. Siegel,
who brought his related work to our attention \cite{Roiban:2000yy} \cite{Dai:2009zg} \cite{Siegel:2010yd}.}
\begin{equation}\label{Gammadef}
\G = I(\th) {\hat f}^{-1} =  {\hat f}^{-1} I(\th),
\end{equation}
where
\begin{equation}
I(\th) = \left( \begin{array}{cc}
I &  \z \th  \\
\z \td & I \\
\end{array} \right)
\end{equation}
and
\begin{equation}
\hat f = \left( \begin{array}{cc}
f &  0  \\
0 & {\tilde f} \\
\end{array} \right).
\end{equation}
In this formula $f$ denotes a real analytic function of the
hermitian matrix $u=i\th\td$ and $\tilde f$
denotes the same function with argument $\tilde u = i\td\th$. These functions are actually
polynomials of degree 16 or less, since higher powers necessarily vanish. The
fact that $[I(\th), \hat f]=0$ is a consequence of the identities
\begin{equation}
f \th = \th \tilde f \quad {\rm and} \quad \td f = \tilde f \td .
\end{equation}

The choice of the function $f$ is determined by requiring that $\G$ is
superunitary, \ie $\G^{\dagger} \G = I$. Using the definition of the
superadjoint given in Eq.~(\ref{superadjoint}),
\begin{equation}\label{Gammainv}
\G^{\dagger} = I(-\th) {\hat f}^{-1} = {\hat f}^{-1} I(-\th).
\end{equation}
The requirement $\G^{\dagger} \G = I$ then becomes
\begin{equation}
{\hat f}^2 = I(-\th) I(\th) = \left( \begin{array}{cc}
I + u & 0  \\
0 & I + \tilde u\\
\end{array} \right) .
\end{equation}
Therefore the correct choices for $f$ and $\tilde f$ are the hermitian matrices
\begin{equation}
f = \sqrt{I+u} = I + \half u + \ldots \quad {\rm and} \quad
\tilde f = \sqrt{I + \tilde u} = I + \half \tilde u +\ldots.
\end{equation}
For this choice $\G$ is an element of the supergroup $SU(2,2|4)$.

\subsection{Grassmann-valued connections}

Various one-forms that can be regarded as connections associated to the superalgebra
will arise in the course of this work. Here we utilize the nonlinear realization that
we just found to construct ones that only involve the Grassmann coordinates.
The $y$ and $z$ coordinates will need to be incorporated later, and then additional connections
will be defined.

Consider the super-Lie-algebra-valued one-form
\begin{equation} \label{Adef}
\cA = \G^{-1} d\G
= \left( \begin{array}{cc}
K &  \z \Psi  \\
\z \Psi^{\dagger} & \tilde K \\
\end{array} \right) .
\end{equation}
This supermatrix is super-antihermitian, as required. (The matrices required to take
account of the indefinite signature of $\mathfrak{su}(2,2)$ are implicit, as discussed
earlier.) Explicit calculation gives
\begin{equation} \label{Kresult}
K = - d f f^{-1} +i\th\Psi^{\dagger} = f^{-1} df -i \Psi \td ,
\end{equation}
and
\begin{equation}
\tilde K = - d \tilde f {\tilde f}^{-1} +i\td\Psi
= {\tilde f}^{-1} d \tilde f -i \Psi^{\dagger} \th ,
\end{equation}
where
\begin{equation} \label{Psidef}
\Psi = f^{-1} d \th {\tilde f}^{-1}
\quad {\rm and} \quad
\Psi^{\dagger} = {\tilde f}^{-1} d \td f^{-1}.
\end{equation}
We prefer to not subtract the trace parts of
$K$ and $\tilde K$, which would be required to make them elements of $\mathfrak{su}(4)$
and $\mathfrak{su}(2,2)$, respectively.
Since we define $\mathfrak{psu}(2,2|4)$ as a quotient space of $\mathfrak{su}(2,2|4)$, it
is sufficient for our purposes that
$\tr K = \tr \tilde K$, which implies that $\Str \cA =0$.
This ensures that the traces could be removed, as in the case of $\o_{12}\th
- \th \tilde\o_{12}$, which was discussed earlier.

The fact that $\cA$ is ``pure gauge'' implies that it is a flat connection, \ie
\begin{equation}
d \cA + \cA \wedge \cA = 0.
\end{equation}
In terms of $4 \times 4$ blocks the zero-curvature equations are
\begin{equation} \label{FMC}
d K + K \wedge K - i\Psi \wedge \Psi^{\dagger} = 0,
\quad
d \tilde K + \tilde K \wedge \tilde K - i\Psi^{\dagger} \wedge \Psi = 0,
\end{equation}
\begin{equation} \label{DPsi}
d \Psi + K \wedge \Psi + \Psi \wedge \tilde K = 0,
\quad
d \Psi^{\dagger} + \tilde K \wedge \Psi^{\dagger} + \Psi^{\dagger} \wedge K = 0.
\end{equation}
These equations have the same structure as the Maurer--Cartan equations of
the superalgebra.

Under an arbitrary variation $\d \G$, the variation of $\cA = \G^{-1} d\G$ is
\begin{equation} \label{deltacA}
\d\cA = d(\G^{-1} \d \G) + [ \cA, \G^{-1} \d \G].
\end{equation}
The supermatrix $\G$ depends only on $\th$. Therefore to determine how $\cA$
varies under an arbitrary variation of $\th$, we need to know the variation $\G^{-1} \d\G$
for an arbitrary variation of $\th$. The result is that
\begin{equation} \label{calMhat}
\G^{-1} \d \G = \left( \begin{array}{cc}
{\cal M} &  \z \r  \\
\z \r^{\dagger} & \tilde{\cal M} \\
\end{array} \right) = \hat{\cal M}
\end{equation}
where
\begin{equation} \label{calM}
{\cal M}  = -\d f f^{-1} + i \th\rd  =  f^{-1} \d f - i \r\td
\end{equation}
\begin{equation} \label{caltM}
\tilde{\cal M}  = -\d \tilde f {\tilde f}^{-1} + i \td \r
=  {\tilde f}^{-1} \d \tilde f - i \rd\th,
\end{equation}
and
\begin{equation} \label{rhodef}
\r = f^{-1} \d \th {\tilde f}^{-1}.
\end{equation}
These formulas for $\G^{-1} \d\G$ have exactly the same structure as the previous ones
for $\cA= \G^{-1} d \G$. So no additional computation was required to
derive them. They will be useful later when we derive equations of motion.

A special case of these formulas that is of particular interest is when the variation
$\d\th$ is not arbitrary but rather is an infinitesimal supersymmetry transformation
of the form given in Eq.~(\ref{deltath}). In that case we find
\begin{equation} \label{deltaGamma}
\d_{\e}\G = \G \left( \begin{array}{cc}
M &  0  \\
0 & \tilde M \\
\end{array} \right) + \left( \begin{array}{cc}
0 &  \z \e  \\
\z \ed & 0 \\
\end{array} \right)\G,
\end{equation}
where
\begin{equation}\label{Mformula}
M = -(\d_{\e} f - i f \e \td) f^{-1} =  f^{-1}( \d_{\e} f - i \th \ed f)
\end{equation}
and
\begin{equation}
\tilde M = -(\d_{\e} \tilde f - i \tilde f \ed \th) \tilde f^{-1}
= \tilde f^{-1}( \d_{\e} \tilde f - i \td \e \tilde f) .
\end{equation}

Equation~(\ref{deltaGamma})
shows that under a supersymmetry transformation $\G$ is multiplied on the right by
a local $\mathfrak{su}(4) \times \mathfrak{su}(2,2)$ transformation and on the left by
a global supersymmetry transformation, just as one would expect in a coset construction.
This supports interpreting the nonlinear realization of $PSU(2,2|4)$
in terms of $\th$ as a coset construction\footnote{This was pointed out by E. Witten.}
\begin{equation}
PSU(2,2|4)/ SU(4) \times SU(2,2).
\end{equation}
The global supersymmetry term, \ie the second term on the right-hand side of
Eq.~(\ref{deltaGamma}), does not contribute to $\d_\e \cA$, which takes the form
\begin{equation}
\d_\e\cA = d(\G^{-1} \d_\e \G) + [ \cA, \G^{-1} \d_\e \G] = d \hat M + [ \cA, \hat M ].
\end{equation}
where
\begin{equation}
\hat M = \left( \begin{array}{cc}
M &  0  \\
0 & \tilde M \\
\end{array} \right).
\end{equation}
In terms of blocks this gives
\begin{equation} \label{deltaK}
\d_{\e} K =  dM + [K,M]\quad {\rm and} \quad
\d_{\e} \tilde K =  d \tilde M +[\tilde K , \tilde M],
\end{equation}
\begin{equation} \label{Psitrans}
\d_{\e} \Psi =  \Psi \tilde M - M\Psi \quad {\rm and} \quad
\d_{\e} \Psi^{\dagger} = \Psi^{\dagger} M - \tilde M\Psi^{\dagger}  .
\end{equation}

\subsection{Inclusion of bosonic coordinates}

The formulas that have been described in this section so far describe the supermanifold
geometry for fixed values of the bosonic coordinates $y$ and $z$, introduced in Sect.~2,
which we sometimes refer to collectively as $x$. Our goal now is to describe the
generalization that also allows the bosonic coordinates to vary. For this purpose,
the first step is to recast $y$ and $z$ as $4 \times 4$ matrices
denoted $Y$ and $Z$. This is described in detail in Appendix~A. The result that is
established there is that
\begin{equation}
Y^{\m\n} = y^m (\tilde\S_m)^{\m\n} \quad {\rm and} \quad
Z^{\a\b} = z^a (\S_a)^{\a\b}
\end{equation}
are antisymmetric matrices belonging to the groups $SU(2,2)$ and $SU(4)$,
respectively. Thus, in our notation, $Y^T = -Y$, $Z^T = -Z$, $\Yd = Y^{\dagger} $,
$ \Zd = Z^{\dagger}$,
and $\det Y = \det Z = 1$. These equations are consequences of the relations
$y^2 = -1$ and $z^2 = 1$, as well as Clifford-algebra-like formulas for the $\S$ matrices.
Thus, $S^5$ is described as a specific codimension 10 submanifold of the $SU(4)$
group manifold, and $AdS_5$ is described as a specific codimension 10 submanifold of the
$SU(2,2)$ group manifold.

The supersymmetry transformations of the bosonic coordinates, which are encoded in the
antisymmetric matrices $Z$ and $Y$, are given by induced local $SU(4)$ and $SU(2,2)$
transformations
\begin{equation} \label{deltaZ}
\d_\e Z = -(M Z + Z M^T)   \quad {\rm and} \quad \d_\e Y = -(\tilde M Y + Y {\tilde M}^T).
\end{equation}
Defining the supermatrix
\begin{equation} \label{Xdef}
X
= \left( \begin{array}{cc}
Z & 0  \\
0 & Y \\
\end{array} \right) ,
\end{equation}
these can be combined into the supermatrix equation
\begin{equation} \label{deltaMX}
\d_\e X = -(\hat M X + X {\hat M}^T) .
\end{equation}
As a check of these formulas, one can verify that the commutator
of two such transformations gives the correct infinitesimal $\mathfrak{su}(4)$ and
$\mathfrak{su}(2,2)$ transformations. This is achieved in the $\mathfrak{su}(4)$ case
by verifying that
\begin{equation} \label{M1M2t}
[M_2,M_1] + \d_2 M_1 - \d_1 M_2 = \o_{12},
\end{equation}
where $M_i = M(\e_i)$, $\d_i = \d_{\e_i}$, and $\o_{12}$ is given in Eq.~(\ref{omegaS}).
The $\d_\e Y$ equation is established in the same way.

\subsection{Majorana--Weyl matrices and Maurer--Cartan equations}

In the flat spacetime limit, a fermionic matrix such as $\th$ corresponds to a complex
Weyl spinor, which (in the notation of \cite{Becker:2007zj})
satisfies an equation of the form $\G_{11} \th = \th$.
This spinor describes a reducible representation of the ${\cal N} = 2B$, $D=10$
super-Poincar\'e group,
and so it can be decomposed into a pair of Majorana--Weyl spinors $\th = \th_1 + i\th_2$.
In a Majorana representation of the Dirac algebra the MW spinors $\th_1$ and
$\th_2$ each contain 16 real components. In the case of $PSU(2,2|4)$ the group theory
is different. The relevant representation of $SU(2,2) \times SU(4)$ is still reducible,
$({\bf 4, \bar 4}) + ({\bf \bar 4, 4})$, but it does not make group-theoretic sense to
extract the real and imaginary parts by adding and subtracting these two pieces.
Fortunately, there is a construction that is group theoretically sensible and connects
smoothly with the flat-space limit.

The transformations given previously imply that $\Psi$ and
\begin{equation} \label{Psiprime}
\Psi' = Z \Psi^{\star} Y^{-1}
\end{equation}
transform in the same way under all $PSU(2,2|4)$ transformations. To understand
this definition one should follow the indices. The complex conjugate is
$(\Psi^\a{}_\m)^{\star} = (\Psi^{\star})^{\bar\a}{}_{\bar\m}$, but as usual we convert
to unbarred indices, $(\Psi^{\star})_{\a}{}^{\m}$,  using $\eta$ matrices, \ie
$\Psi^{\star} \to \eta\Psi^{\star} \eta$. Then $(\Psi')^\a{}_\m = Z^{\a\b}
(\Psi^{\star})_\b{}^\n (\Yd)_{\n\m}$.  Therefore $\Psi$ and $\Psi'$ transform in the
same way, and it makes group-theoretic sense to define
\begin{equation}
\Psi_1 = \half(\Psi + \Psi') \quad {\rm and} \quad \Psi _2 = \frac{1}{2i}(\Psi - \Psi').
\end{equation}
Then $\Psi = \Psi_1 + i \Psi_2$ and $\Psi' = \Psi_1 -i \Psi_2$. We will
refer to $\Psi_1$ and $\Psi_2$ as {\em Majorana--Weyl matrices}. A MW matrix,
such as $\Psi_1$, satisfies the ``reality'' identity
\begin{equation}
\Psi_1 =  Z\Psi_1^{\star} \Yd \quad {\rm or} \quad  \Psi_1^{\dagger} = Y \Psi_1^T \Zd.
\end{equation}

What we have here is a generalization of complex conjugation given by
\begin{equation} \label{rhoprime}
\r \to \m(\r) = \r' = Z \r^{\star} \Yd,
\end{equation}
where $\r^\a {}_\m$ is an arbitrary fermionic matrix (not necessarily a one-form)
that transforms under $SU(2,2) \times SU(4)$ transformations like $\Psi$ or $\th$.
Using the antisymmetry and unitarity of $Y$ and $Z$ it is easy to verify that
$\m$ is an involution, like complex conjugation, \ie
$\m \circ \m = I$, where $I$ is the identity operator. Therefore,
\begin{equation}
\m_\pm = \half(I \pm \m)
\end{equation}
are a pair of orthogonal projection operators that separate $\r$ into two
pieces, $\r = \r_1 + i \r_2$. In the flat-space limit $\r_1$ and $\r_2$
correspond to conventional MW spinors.

There are two possible definitions of the covariant exterior derivative of $X$, the
supermatrix that represents the bosonic coordinates $y$ and $z$. They are
\begin{equation}
D_+ X = dX + {\cal A} X + X {\cal A}^T \quad {\rm and} \quad
D_- X = dX + {\cal A} X + X {\cal A}^{\overline T} .
\end{equation}
Here, $\overline T$ denotes the inverse transpose, which acts on the odd blocks with the opposite sign
from the transpose $T$. It may be surprising that
$D_+ X$ and $D_- X$ have nonvanishing odd blocks, even though $X$ does not. There is no inconsistency,
and these are definitely the most convenient and natural definitions. In particular,
$D_{\pm} X$ transform under a supersymmetry transformation in the same way as $X$, namely
\begin{equation} \label{deltaDX}
\d_{\e} D_{\pm} X = -(\hat M D_{\pm} X  + D_{\pm} X{\hat M}^T).
\end{equation}
Note that ${\hat M}^T = {\hat M}^{\overline T}$, since ${\hat M}$ only has even blocks.
The antisymmetry of $X$ gives rise to the relation $(D_- X)^T = - D_+ X$.

Given these definitions, we can define a pair of connections
\begin{equation}
A_+ =  - D_+ X X^{-1} = - dX X^{-1} - {\cal A} - X {\cal A}^T X^{-1}
\end{equation}
and
\begin{equation}
A_- =  - D_- X X^{-1} = - dX X^{-1} - {\cal A} - X {\cal A}^{\overline T} X^{-1}.
\end{equation}
Inserting the definition of ${\cal A}$ and remembering the factors of $\pm i$ in the
odd blocks of ${\cal A}^T$ and ${\cal A}^{\overline T}$, these connections
can be written in the form
\begin{equation}
A_{\pm} =  A_1 + A_2 \pm i A_3,
\end{equation}
where
\begin{equation}
A_1
= \left( \begin{array}{cc}
\O & 0  \\
0 & \tilde\O \\
\end{array} \right) ,
\quad
A_2
= -\left( \begin{array}{cc}
0 &  \z \Psi  \\
\z \Psi^{\dagger} & 0 \\
\end{array} \right) ,
\quad
A_3
= \left( \begin{array}{cc}
0 &  \z \Psi'  \\
\z \Psi^{\prime\dagger} & 0 \\
\end{array} \right) .
\end{equation}
Also,
\begin{equation} \label{Omegaforms}
{\O}  = Z d \Zd -  K - Z { K}^T \Zd
\quad {\rm and} \quad
{\tilde\O} = Y d \Yd - \tilde K - Y {\tilde K}^T \Yd.
\end{equation}
Each of the three superconnections $A_i$ is super antihermitian, and
under a supersymmetry transformation
\begin{equation}
\d_{\e} A_i = [A_i, \hat M] \quad i=1,2,3.
\end{equation}

Using the definition of the supertranspose in Eq.~(\ref{supertranspose}),
these supermatrices have the important properties
\begin{equation} \label{Atrans}
X A_1^T X^{-1} = A_1, \quad X A_2^T X^{-1} = i A_3, \quad X A_3^T X^{-1} = i A_2 .
\end{equation}
and therefore
\begin{equation} \label{Aplus}
X A_-^T X^{-1} = A_+ \quad {\rm and} \quad X A_+^{\overline T} X^{-1} = A_- .
\end{equation}

Another useful set of one-form supermatrices is
\begin{equation} \label{Jdef}
J_i = \G A_i \G^{-1} ,
\end{equation}
where $\G$ is defined in Eq.~(\ref{Gammadef}). For example,
\begin{equation} \label{J1}
J_1 = \left( \begin{array}{cc}
f^{-1}(\O +i \th \tilde\O \td)f^{-1}& \z f^{-1}(\O\th - \th \tilde\O){\tilde f}^{-1} \\
\z {\tilde f}^{-1}(\tilde\O \td - \td \O )f^{-1}
& {\tilde f}^{-1}(\tilde\O +i \td \O \th){\tilde f}^{-1} \\
\end{array} \right).
\end{equation}
Note that
\begin{equation}
D A_i = dA_i + {\cal A}\wedge A_i + A_i \wedge {\cal A} = \G^{-1} (dJ_i) \G .
\end{equation}
Since $A_i$ is antihermitian, the definition of $D A_i$ does not involve transposes,
and it is unambiguous. Utilizing Eq.~(\ref{deltaGamma}), one can show that
the transformation of these supermatrices under arbitrary
infinitesimal $\mathfrak{psu}(2,2|4)$ transformations is given by \footnote{ Since $\L$
and $\L + i cI$, where $I$ is a unit supermatrix, give the same transformation, the formula only
depends on the equivalence class of $\mathfrak{su}(2,2|4)$ matrices that define a
$\mathfrak{psu}(2,2|4)$ element.}
\begin{equation}
\d_{\L} J_i = [ \L , J_i ],
\end{equation}
where the various infinitesimal parameters are combined in the supermatrix
\begin{equation} \label{Lambdadef}
\L = \left( \begin{array}{cc}
\o &  \z \e  \\
\z \e^{\dagger} & \tilde\o \\
\end{array} \right) .
\end{equation}
The local generalization of these global symmetry transformation rules are used in Appendix~B
to derive the $\mathfrak{psu}(2,2|4)$ Noether currents of the superstring.

The one-forms $J_{\pm} = J_1 + J_2 \pm i J_3$ can be recast in the form
\begin{equation}
J_+ = B_+ d B_+^{-1} \quad {\rm and} \quad J_- = B_- d B_-^{-1},
\end{equation}
where
\begin{equation}\label{Bpm}
B_+ = \G X \G^T \quad {\rm and} \quad B_- = \G X \G^{\overline T}.
\end{equation}
These relations imply that $J_+$ and $J_-$ are flat connections
\begin{equation}
dJ_+ + J_+ \wedge J_+ = 0 \quad {\rm and} \quad dJ_- + J_- \wedge J_- = 0.
\end{equation}
It is straightforward to verify that $DA_2 = - 2 A_2 \wedge A_2$, which implies
that $2J_2$ is also a flat connection. The three flatness conditions
imply the Maurer--Cartan (MC) equations
\begin{equation} \label{dJ1}
dJ_1 = -J_1 \wedge J_1 + J_2 \wedge J_2
+J_3 \wedge J_3 -J_1 \wedge J_2 -J_2 \wedge J_1,
\end{equation}
\begin{equation} \label{dJ2}
dJ_2 = - 2 J_2 \wedge J_2,
\end{equation}
\begin{equation} \label{dJ3}
dJ_3 = -(J_1 + J_2)\wedge J_3
-J_3 \wedge (J_1 + J_2).
\end{equation}

\subsection{Invariant differential forms}

Let us consider the construction of differential forms that are closed and
invariant under the entire supergroup. Such differential forms of degree $p+2$ are required
to construct the Wess--Zumino (WZ) terms of $p$-brane actions. Type IIB superstring
theory has an infinite $SL(2,\IZ)$ multiplet of $(p,q)$ strings, but we
are primarily interested in the fundamental $(1,0)$ superstring here. Its
WZ term is determined by a three-form. Similarly, the D3-brane world-volume action
contains a WZ term determined by a closed and invariant self-dual five-form.

Consider $T_n = \Str( J \wedge J \ldots \wedge J)$, the supertrace of
an $n$-fold wedge product of a one-form supermatrix $J$. This vanishes
for $n$ even because the cyclic identity
of the supertrace, $\Str (AB) = \Str(BA)$, acquires an additional minus sign, \ie
$\Str (A \wedge B) = - \Str(B \wedge A)$,
if $A$ and $B$ are supermatrices of differential forms of odd degree. Now
suppose that $n$ is odd, so that $T_n$ can be nonzero, and that
$J$ is a flat connection ($dJ = - J \wedge J$). In this case the exterior derivative
$d\, T_n$ is proportional to $T_{n+1}$, which is equal to zero. Therefore
$T_n$ is closed.

Let us now utilize this logic to construct a closed three-form based on the
flat connections that we have found. The simple choice $\Str(J_2 \wedge
J_2 \wedge J_2 )= \Str(A_2 \wedge A_2 \wedge A_2 )$ is closed, but it is also zero,
since the product of the three $A_2$ factors has vanishing blocks on the diagonal.
Therefore, let us consider instead
\begin{equation}
T_3 = \Str(J_+ \wedge J_+ \wedge J_+ ).
\end{equation}
This is complex, and therefore it appears to encode two real three-forms that are $PSU(2,2|4)$
invariant and closed. However, Eqs.~(\ref{Aplus}) and (\ref{Jdef}) imply
that $T_3 = - \Str(J_- \wedge J_- \wedge J_- )$, and therefore
the real part of $T_3$ vanishes.

Now let us substitute $J_1 + J_2 + i J_3$ for $J_+$. Doing this,
and only keeping those terms that can give a nonzero
contribution to the supertrace, leaves $T_3 = 3 i T_F$,
where
\begin{equation}\label{TFeqn}
T_F = \Str(J_1 \wedge [J_2 \wedge J_3
 + J_3 \wedge J_2] ).
\end{equation}
The notation is meant to indicate that $T_F$ enters in the
construction of the WZ term of the fundamental string.
The closed three-form $T_F$ is exact, since
\begin{equation}
T_F =  d \, \Str(J_2 \wedge J_3) .
\end{equation}
Therefore the WZ term of the fundamental superstring world-sheet action is proportional
to $\int \Str(J_2 \wedge J_3)$.

The WZ term derived above can be obtained more directly if one anticipates
that the three form is exact. Consider
all invariant two-forms of the type $T_{ij} =\Str(J_i \wedge J_j)$.
Since $J_i$ is a one-form, cyclic permutation gives $T_{ij} = - T_{ji}$.
Furthermore, $T_{13} = T_{23}=0$, because the expressions inside these supertraces
contain no diagonal blocks when reexpressed in terms of $A$'s.
Thus, up to normalization, the only nonzero invariant two-form
of this type is $T_{23}$, which is the one required to construct the
WZ term for the fundamental superstring.

A self-dual five-form, which is closed and $PSU(2,2|4)$ invariant,
also plays an important role in type IIB superstring theory
in an $AdS_5 \times S^5$ background. It has a nonzero bosonic truncation
in contrast to the three-form described above. Its bosonic part is
proportional to the sum (or difference) of the volume form of $AdS_5$
and the volume form of $S^5$, and therefore it is not exact. The supersymmetric
completion of this five-form is proportional to
\begin{equation}
T_5 = \Str(J_+ \wedge J_+ \wedge J_+ \wedge J_+ \wedge J_+ )
= \Str(J_- \wedge J_- \wedge J_- \wedge J_- \wedge J_- ).
\end{equation}
This five-form determines the WZ term for the D3-brane
world-volume action in the $AdS_5 \times S^5$ background.
The complete construction of the D3-brane action will be
described elsewhere.

\section{The superstring world-sheet theory}

\subsection{The action}

The world-volume actions of supersymmetric probe branes, including the fundamental superstring,
are written as a sum of two terms.
The first term, which we denote $S_1$, is of the Nambu--Goto/Volkov--Akulov
type.\footnote{Nambu--Goto refers to a pull-back of the target-space metric to the brane.
Volkov--Akulov refers to the appearance of Goldstino fields in theories with
spontaneously broken supersymmetries -- conformal supersymmetries in the present case.
In the case of D-branes, the $S_1$ term also contains a $U(1)$ field strength and
is usually said to be of the Dirac--Born--Infeld (DBI) type.}
The second term, which we denote $S_2$, is of the Wess--Zumino (WZ) or Chern--Simons type.
Each of these terms is required to have local reparametrization invariance. In the
case of $S_1$, this symmetry can be implemented by introducing a world-sheet metric,
as described in Sect.~2. The $S_2$ term, on the other hand, is independent
of the world-sheet metric. The
target superspace isometry, which in the present case is $PSU(2,2|4)$,
is realized as a global symmetry of $S_1$ and $S_2$ separately.
Furthermore, there should be a local fermionic symmetry, called kappa symmetry.
Kappa symmetry implies that half of the target-space Grassmann coordinates $\th$
are redundant gauge degrees of freedom of the world-volume theory that can be eliminated
by a suitable gauge choice. Unlike all of the other symmetries, kappa symmetry
is not a symmetry of $S_1$ and $S_2$ separately. Rather, it requires a conspiracy between
them. Given $S_1$, a specific $S_2$, unique up to sign,
is required by kappa symmetry. In the case of flat ten-dimensional spacetime,
the superstring action $S = S_1 + S_2$ turns out to be a free theory,
a fact that can be made manifest in light-cone gauge \cite{Green:1983wt}.
In this gauge the exact spectrum of the string
is easily determined. In the case of an $AdS_5 \times S^5$ background, the superstring
world-sheet theory is not a free theory, but it is an integrable theory \cite{Bena:2003wd},
as we will discuss later.

The construction of $S_1$ works exactly as explained for the bosonic truncation
in Sect.~2. The only change is that now $G_{\a\b}$ is determined by
a supersymmetrized target-space metric. The correct choice turns out to be
\begin{equation} \label{Gform}
G_{\a\b}
= -\frac{1}{4}\Str(J_{1\a} J_{1\b}) = -\frac{1}{4}\Str(A_{1\a} A_{1\b})
= -\frac{1}{4} \left(\tr(\O_\a \O_\b) - \tr(\tilde\O_\a\tilde\O_\b) \right).
\end{equation}
The normalization is chosen to give the correct bosonic truncation.
As explained in Sect.~2,
\begin{equation}
S_1 = - \frac{\sqrt{\l}}{4\pi}\int d^2 \s \sqrt{-h} h^{\a\b}G_{\a\b} ,
\end{equation}
which is classically equivalent to
\begin{equation}
S_1 = - \frac{\sqrt{\l}}{2\pi}\int d^2 \s \sqrt{- \det G_{\a\b} }.
\end{equation}
%A priori, $G_{\a\b}$ could also contain a term proportional to
%$\Str(J_{2\a} J_{2\b})= -\Str(J_{3\a} J_{3\b})$. This would preserve all
%of the global symmetry and not affect the bosonic truncation. However, it turns out that
%kappa symmetry does not allow such a contribution.

The second term in the superstring world-sheet action, denoted $S_2$, should
also be invariant under the entire $PSU(2,2|4)$ supergroup. Furthermore,
as discussed in Sect.~3.6, it must be proportional to
$\int \Str(J_2 \wedge J_3)$. Its normalization should be chosen such that
$S = S_1 + S_2$ has local kappa symmetry. This symmetry implies
that half of the $\th$ coordinates are gauge degrees of freedom of the world-sheet
theory. Altogether, the superstring action is
\begin{equation}
S =  \frac{\sqrt{\l}}{16\pi}\int d^2 \s \left( \sqrt{-h} h^{\a\b}\Str(J_{1\a} J_{1\b})
+ k \, \e^{\a\b}\Str(J_{2\a} J_{3\b}) \right),
\end{equation}
where the coefficient $k$ will be determined by requiring kappa symmetry, though its
sign is a matter of convention. For any $k$
this action has manifest global $PSU(2,2|4)$ symmetry, since for an arbitrary
infinitesimal transformation by a constant amount $\L$, $\d_\L J_i = [\L,J_i]$.
We will find that kappa symmetry requires $k = \pm 2$ and choose the plus sign.
Then the action can be rewritten in the form
\begin{equation} \label{action}
S =  \frac{\sqrt{\l}}{16\pi}\int \left(\, \Str(J_1\wedge \star J_1)
+ 2\, \Str(J_2 \wedge J_3) \right).
\end{equation}
The Hodge dual is defined here using the metric $h_{\a\b}$, though (at leading order in the
world sheet genus expansion) it is equivalent to replace $h_{\a\b}$ by $G_{\a\b}$.

\subsection{Equations of motion}

There are conserved currents, called Noether currents, associated to
each of the generators of the super Lie algebra $\mathfrak{psu}(2,2|4)$.
The derivation of these Noether currents is given in Appendix~B. The result
obtained there is
\begin{equation} \label{JN}
J_N = J_1 + \star J_3.
\end{equation}
The statement that these currents are conserved is
\begin{equation} \label{dJN}
d \star J_N = d( \star J_1 + J_3 ) =0.
\end{equation}
These conservation equations hold as a consequence of the equations of motion.
In fact, as in the case of the bosonic truncation, they are equivalent to the equations of
motion that are obtained from arbitrary variations  $\d X$ and $\d\th$.
Note that the Noether current $J_1 + \star J_3$ is not flat, even though
(as noted earlier) the Noether current for the bosonic truncation of the theory is flat.

In order to set the stage for the later proof of kappa symmetry, let us consider
arbitrary variations of the Grassmann coordinates
$\d \th $ (and $\d \td $). Using Eqs.~(\ref{deltacA})--(\ref{rhodef}),
this determines the variation of $\cA$, which depends only on $\th$, to be
\begin{equation} \label{deltakapA}
\d \cA = D \hat{\cal M} = d \hat{\cal M} + [\cA, \hat{\cal M}],
\end{equation}
where
\begin{equation}
\hat{\cal M} = \G^{-1} \d \G = {\cal N} + R,
\end{equation}
\begin{equation} \label{Ndef}
{\cal N} = (\hat{\cal M})_{\rm even}
= \left( \begin{array}{cc}
{\cal M} & 0  \\
0 & \tilde{\cal M} \\
\end{array} \right),
\end{equation}
\begin{equation} \label{Rdef}
R = (\hat{\cal M})_{\rm odd}
= \left( \begin{array}{cc}
0 & \z \r  \\
\z \r^{\dagger} & 0 \\
\end{array} \right).
\end{equation}
Let us simultaneously vary the bosonic coordinates, which are encoded in $X$.
We make a specific choice that is required to reveal local kappa symmetry, namely
\begin{equation} \label{deltaX}
\d X = -({\cal N} X + X {\cal N}^T).
\end{equation}
Like $X$, the right-hand side is even and antisymmetric. Also,
${\cal N}^T = {\cal N}^{\overline T}$ is unambiguous.

Decomposing ${\cal A}$ into even and odd parts,
\begin{equation}
{\cal A} = \hat K - A_2,
\end{equation}
Eq.~(\ref{deltakapA}) decomposes into the pair of equations
\begin{equation}
\d \hat K = d {\cal N}  + [\hat K, {\cal N}]  + \hat\D,
\end{equation}
and
\begin{equation} \label{deltaA2}
\d A_2 =  - (D R)_{\rm odd} + [ A_2, {\cal N}],
\end{equation}
where
\begin{equation}
\hat\D  = (DR)_{\rm even} = - [A_2 , R].
%= \left( \begin{array}{cc}
%\D & 0  \\
%0 & \tilde\D \\
%\end{array} \right) .
\end{equation}

Let us now evaluate the variation of $S_1$, the first term in the Lagrangian, which
is proportional to the integral of
\begin{equation}
\Str(J_1 \wedge \star J_1) = \Str(A_1 \wedge \star A_1) .
\end{equation}
To compute the variation of this expression, we require the variation of
\begin{equation}
A_1 = XdX^{-1} - \hat K - X {\hat K}^T X^{-1}.
\end{equation}
Using the equations given above, one obtains
\begin{equation}\label{dRA1}
\d A_1 = [A_1,{\cal N}] - \hat\D - X {\hat\D}^T X^{-1}.
\end{equation}
Therefore,
\begin{equation}
\d \, \Str (A_1 \wedge \star A_1) = -2 \,\Str((\hat\D +X {\hat\D}^T X^{-1}) \wedge \star A_1)
 = -4 \,\Str(\hat\D \wedge \star A_1) .
\end{equation}
This gives
\begin{equation} \label{deltads2}
\d \, \Str (A_1 \wedge \star A_1) = 4 \,\Str([A_2 , R] \wedge \star A_1)
=  4\, \Str(R[A_1 \wedge \star A_2 + \star A_2 \wedge A_1]).
\end{equation}
In the last step we have used the identity $ \star A_1 \wedge A_2 + A_1 \wedge \star A_2 =0$.
Altogether,
\begin{equation} \label{deltaS1a}
\d S_1 = \frac{\sqrt{\l}}{4\pi} \int \Str(R[A_1 \wedge \star A_2 + \star A_2 \wedge A_1]).
\end{equation}

Turning to the WZ term, $S_2$, we need to compute the variation of
\begin{equation}
\Str(J_2 \wedge J_3) = \Str(A_2 \wedge A_3) = -i \, \Str(A_2 \wedge X A_2^T X^{-1}).
\end{equation}
Using Eq.~(\ref{deltaA2}),
\begin{equation}
\d \, \Str(J_2 \wedge J_3) = 2\, \Str(\d A_2 \wedge A_3) = -2\, \Str (D R \wedge A_3),
\end{equation}
where we put back the even part of $DR$, since it does not contribute to the supertrace.
The variation $\d S_1$ does not involve any derivatives of $R$, but the expression
we have just found does contain one. Thus, if these two terms are to combine nicely, an
integration by parts is required. The appropriate formula is
\begin{equation}
\Str(D R \wedge A_3) = d \, \Str(R A_3) - \Str(R D A_3).
\end{equation}
Applying the general rule $D A_i = \G^{-1} d J_i \G$
to $DA_3$, and using the MC equation for $dJ_3$, we find
\begin{equation}
(D A_3)_{\rm odd} =  -(A_3 \wedge A_1 + A_1 \wedge A_3),
\end{equation}
and therefore
\begin{equation} \label{delta23}
\d\, \Str(J_2 \wedge J_3)
= - 2 \, \Str(R [A_3 \wedge A_1 + A_1 \wedge A_3]) - 2d \, \Str(R A_3).
\end{equation}
Adjusting the normalization of the WZ term by setting
$k=2$, and dropping the total differential, gives the variation
\begin{equation} \label{deltaS2}
\d S_2  = - \frac{\sqrt{\l}}{4\pi}\int \Str(R [A_3 \wedge A_1 + A_1 \wedge A_3]).
\end{equation}

Combining Eqs.~(\ref{deltaS1a}) and (\ref{deltaS2}),
\begin{equation} \label{deltaS}
\d S  = \frac{\sqrt{\l}}{4\pi}\int
\Str(R [(\star A_2 - A_3) \wedge A_1 + A_1 \wedge (\star A_2 - A_3)]).
\end{equation}
This implies that
\begin{equation} \label{2ddual}
(\star A_2 - A_3) \wedge A_1 + A_1 \wedge (\star A_2 - A_3) =0
\end{equation}
is an equation of motion. Equivalently,
\begin{equation} \label{2ddualJ}
(\star J_2 - J_3) \wedge J_1 + J_1 \wedge (\star J_2 - J_3) =0.
\end{equation}

As we discussed earlier, the $h_{\a\b}$ equation of motion implies that $h_{\a\b}$ is proportional to
$G_{\a\b}$. Therefore, even though the Hodge dual was defined using $h_{\a\b}$ in the preceding equations
of motion, it can equivalently be defined using the induced metric $G_{\a\b}$. Had we started
with the $\sqrt {-G}$ form of $S_1$, we would have obtained the same equations of motion,
but with the $G_{\a\b}$ form of the Hodge dual in the first place.

In this section we have analyzed two types of variations. The first
exploited the global symmetry of the theory to construct the corresponding
conserved currents $J_N$, whose conservation encodes equations of motion.
This entailed studying variations in which the infinitesimal $\mathfrak{psu}(2,2|4)$
parameters encoded in the supermatrix $\L$ are functions of the world-sheet
coordinates. The second variation we considered allowed an arbitrary local
variation of the Grassmann coordinates $\d\th$ together with specific
variations of the bosonic coordinates $\d X$ completely determined by $\d\th$.
%This led to the derivation of additional equations of motion, and it will also be
%utilized to prove local kappa symmetry.
In the case of local $\L$ variations we found (in Appendix~B) that the $d\L$ part of the
variations of the one-form supermatrices are concisely encoded in the single formula
\begin{equation}
\d' A_+  = -\G^{-1} d\L \G - X [\G^{-1} d\L \G]^T X^{-1}.
\end{equation}
Using Eqs.~(\ref{deltaA2}) and (\ref{dRA1}), one can deduce an analogous formula
for the second type of variation
\begin{equation}
\d' A_+  = -DR - X [DR]^T X^{-1}.
\end{equation}
Using the identity $DR = \G^{-1} d(\G R\G^{-1})\G$, this suggests the correspondence
\begin{equation}
\L \sim \G R \G^{-1}.
\end{equation}
In other words, a special class of local $\L$ parameters are determined by $\d\th$
of the second type of variation. In terms of components, the correspondence is
\begin{equation}
\e(\s) = f^{-2} ( \d\th +i \th \d\th^{\dagger} \th) {\tilde f}^{-2}
\end{equation}
\begin{equation}
\o(\s) = if^{-2} ( \d\th \th^{\dagger} - \th \d\th^{\dagger} ) {f}^{-2}
\end{equation}
\begin{equation}
\tilde\o(\s) = i{\tilde f}^{-2} ( \d\th^{\dagger} \th - \th^{\dagger} \d\th ) {\tilde f}^{-2}.
\end{equation}
For these choices $R = \G^{-1} \L \G$.
Therefore the equations of motion (\ref{2ddual}) must actually be a special case of
the conservation of the Noether current.
In fact, Eq.~(\ref{2ddual}) is equivalent to the odd part of the Noether current conservation
equation written in the form $D(\star A_1 + A_3) =0$.

\subsection{Integrability}

In general, for two matrices of one-forms $J_1$ and $J_2$ in 2d, with a Lorentzian
signature metric,
\begin{equation}
\star J_1 \wedge J_2 +  J_1 \wedge \star J_2 =0 .
\end{equation}
Using this fact, Eq.~(\ref{2ddualJ}) can be rewritten in the form
\begin{equation}
\star J_1 \wedge J_2 + J_2 \wedge \star J_1  +  J_1 \wedge J_3 + J_3 \wedge J_1 =0.
\end{equation}
By taking the transpose of this equation and conjugating by $X$ one deduces that
%\begin{equation}
%(\star J_3 + J_2) \wedge J_1 + J_1 \wedge (\star J_3 + J_2) =0
%\end{equation}
%or
\begin{equation}
\star J_1 \wedge J_3 + J_3 \wedge \star J_1  +  J_1 \wedge J_2 + J_2 \wedge J_1 =0.
\end{equation}

As has been explained, the last two equations are consequences of the
conservation of the Noether current, $d (\star J_1 + J_3) =0$. This equation
and the MC equations (\ref{dJ1})--(\ref{dJ3}) are the ingredients required for the proof
of integrability given in \cite{Bena:2003wd}. Specifically, in terms of a spectral
parameter $x$, the supermatrix of currents (or connections)
\begin{equation}
J(x) = c_1 J_1 + c_1'\star J_1 + c_2 J_2 + c_3 J_3
\end{equation}
is flat (\ie $dJ + J \wedge J =0$) for
\begin{equation}
c_1 = - \sinh^2 x, \quad c_1' = \pm \sinh x \cosh x, \quad
c_2 = 1 \mp \cosh x, \quad c_3 = \sinh x .
\end{equation}
These equations allow one to construct an infinite family of conserved charges
and establish integrability.
The integrability of this theory was explored further in \cite{Alday:2005gi}.

\subsection{Kappa symmetry}

The variation of the action in Eq.~(\ref{deltaS}) is proportional to $\int W$, where
\begin{equation}
W = \Str(R [C \wedge A_1 + A_1 \wedge C]) = \Str([R, A_1] \wedge C),
\end{equation}
and
\begin{equation}
C= \star A_2 - A_3.
\end{equation}
In this section we will derive the local variations $\d \th$ and $\d X$ for which $W$
vanishes, up to an exact two-form, thereby deriving the local kappa symmetry
transformations.

As discussed earlier, the Hodge dual that appears
in $C$ can be defined using either the auxiliary metric $h_{\a\b}$ or the induced metric
$G_{\a\b}$ depending on which form of the action is used.
This choice does not matter for deriving classical equations of motion, since the $h$
equation of motion relates $h_{\a\b}$ to $G_{\a\b}$. However, kappa symmetry is
supposed to be a symmetry of the action, so equations of motion should not be invoked.
The formulas in Appendix~C, which will enable us to prove kappa symmetry, require using the
induced metric $G_{\a\b}$ to define the Hodge dual.
Therefore, this will be the meaning of the Hodge dual for the remainder
of this section and in Appendix~C.

%It would be important to retain
%an independent world-sheet metric $h_{\a\b}$ if one wished to consider the
%%string loop expansion, which involves higher-genus Riemann surfaces. However, this is not
%required for studies of AdS/CFT duality in the planar/large-$N$ limit.
%Even in the case of flat 10d spacetime,
%this formalism encounters well-known difficulties for higher-genus Riemann surfaces,
%so one shouldn't expect to do better than that.

The supermatrix $C$ has a special property, namely
\begin{equation} \label{Cprime}
C'= iX C^T X^{-1} = \star \, C ,
\end{equation}
since $X A_2^T X^{-1} = i A_3$ and $X A_3^T X^{-1} = i A_2$.
This relationship, which is crucial for the proof of kappa symmetry, works
because we have chosen the coefficient of the WZ term appropriately.

It is convenient is to decompose $R$ and $C$ into MW supermatrices. This means writing
$R=R_1 +i R_2$ and $C = C_1 + i C_2$ such that
\begin{equation} \label{Rprime}
R'= iX R^T X^{-1} = R_1 -iR_2.
\end{equation}
Since $R$ is antihermitian, $R_1' =R_1$ is antihermitian and
$R_2'=-R_2$ is hermitian. The decomposition
of $C$ works in the same way. Then Eq.~(\ref{Cprime}) implies that
\begin{equation}
C_1 = \star \, C_1 \quad {\rm and} \quad C_2 = - \star C_2.
\end{equation}
Substituting these supermatrices, $W$ takes the form
\begin{equation}
W =  \Str( [R_1, A_1] \wedge C_1) -  \Str( [R_2, A_1] \wedge C_2) .
\end{equation}

Next we invoke the identity derived in Appendix~C
\begin{equation} %\label{keyid}
[R_i, A_1] =  [\g(R_i), \star A_1],
\end{equation}
and the duality properties of $C_i$ given above, to deduce that
\begin{equation}
 [R_1, A_1] \wedge C_1 = [\g_-(R_1), A_1] \wedge C_1
\end{equation}
and
\begin{equation}
 [R_2, A_1] \wedge C_2 = [\g_+(R_2), A_1] \wedge C_2 ,
\end{equation}
where we have introduced projection operators $\g_{\pm} = \half \left(I \pm \g\right)$, so that
\begin{equation}
\g_{\pm}(R) = \half \left(R \pm \g(R)\right).
\end{equation}
Since $\g_+ \circ \g_- = \g_- \circ \g_+ =0$, $W$ vanishes and the action is invariant for the choices
\begin{equation}
\r_1 = \g_+ (\k) \quad {\rm and} \quad \r_2 = \g_-(\k),
\end{equation}
where $\k$  is an arbitrary (local) MW matrix. Since $\th$ describes 32 real
fermionic coordinates, this means that half of them are gauge degrees of freedom,
which can be eliminated by a gauge choice. Recalling that $\r = \r_1 + i \r_2
= f^{-1} \d\th \tilde f^{-1}$, we see that under a kappa symmetry transformation
\begin{equation}
\d_\k \th = f \big(\g_+(\k) + i \g_-(\k)\big) \tilde f .
\end{equation}
The bosonic coordinates $Y$ and $Z$ are varied at the same time in the way described
in Eqs.~(\ref{deltaX}), (\ref{Ndef}). (See also Eqs.~(\ref{calM}),
(\ref{caltM}), and (\ref{Xdef}).)

The superspace $(x,\th)$ has $10+32$ dimensions. However, the local
reparametrization and kappa symmetries imply that only $8+16$ of them
induce independent off-shell degrees of freedom of the superstring, just
as in flat 10d spacetime. In the case of the flat spacetime theory, there is a gauge choice
for which the superstring world-sheet theory becomes a free theory \cite{Green:1983wt}.
This is certainly not the case for the $AdS_5 \times S^5$ background, though some
choices are more convenient than others. Possible
gauge choices for the $AdS_5 \times S^5$ theory have been discussed extensively beginning
with \cite{Kallosh:1998zx} \cite{Kallosh:1998qv} \cite{Kallosh:1998nx} \cite{Kallosh:1998ji}.
This important issue will not be pursued here.

\section{Conclusion}

The superspace geometry of the $AdS_5 \times S^5$
solution of type IIB superstring theory and the dynamics of a fundamental
superstring embedded in this geometry have been reexamined from a somewhat new perspective.
We began by presenting a nonlinear realization of the superspace isometry
supergroup $PSU(2,2|4)$ in terms of Grassmann coordinates only. The resulting
formulas were interpreted as arising from a $PSU(2,2|4)/ SU(2,2)\times SU(4)$
coset construction. Following that, unitary antisymmetric matrices $Z= \S \cdot \hat z$
and $Y = \tilde\S \cdot \hat y$ were introduced to describe the $S^5$ and $AdS_5$ coordinates,
respectively. These matrices were interpreted as describing specific embeddings of $S^5$
inside $SU(4)$ and $AdS_5$ inside $SU(2,2)$.

Next we constructed three supermatrix one-forms $J_1$, $J_2$,
and $J_3$ that transform linearly under infinitesimal global
$\mathfrak{psu}(2,2|4)$ transformations,
$\d J_i = [\L, J_i]$. In terms of these one-forms the superstring world-sheet
action was shown to be
\begin{equation}
S =  \frac{\sqrt{\l}}{16\pi}\int \left(\, \Str(J_1\wedge \star J_1)
- 2\, \Str(J_2 \wedge J_3) \right).
\end{equation}
This action has manifest global $\mathfrak{psu}(2,2|4)$
symmetry and manifest local reparametrization invariance.
The Hodge dual in the first term can be defined using either an auxiliary metric $h_{\a\b}$
or the induced metric $G_{\a\b} = - \frac{1}{4} (\Str J_{1\a} J_{1\b})$.
However, the latter choice is required to establish local
kappa symmetry, which is not manifest.  Kappa symmetry
was shown to arise from an interplay of three involutions. It determines
the coefficient of the second term in the action, up to a sign that
is convention dependent.
Conservation of the $\mathfrak{psu}(2,2|4)$ Noether current $J_N = J_1 \, + \, \star J_3$
encodes the equations of motion.
Using these equations, a one-parameter family of flat connections,
required for the proof of integrability, was obtained.
%Some of the $\mathfrak{psu}(2,2|4)$ symmetry is spontaneously broken, and
%all of the degrees of freedom of the superstring action are Goldstone modes.

All of these results are in complete agreement with what others have found long ago.
So far, the main achievement of this work is to reproduce well-known results.
However, the formulation described here has some attractive features that
are not shared by previous ones. For one thing, the complete dependence of
all quantities on the Grassmann coordinates is described by simple analytic
expressions. Also, the action and the equations of motion have
manifest global $PSU(2,2|4)$ symmetry.
In particular, at no point did we need to
decompose $\mathfrak{psu}(2,2|4)$ into pieces,\footnote{The generators associated
to these pieces are usually denoted $P$, $Q$, $D$, $J$, $R$, $S$, $K$.} as is often done.
%The utility of this formalism for obtaining new results remains to be demonstrated.

There are two main directions that we hope to explore in the future
using the results obtained here.
One is to derive new facts about the dynamics of this fundamental superstring.
The other is to explore other brane theories, such as a probe D3-brane embedded
in the same $AdS_5 \times S^5$ background or a fundamental type IIA superstring in an
$AdS_4 \times CP^3$ background.

\section*{Acknowledgments}

The author acknowledges discussions with Matthew Heydeman as well as helpful
comments from Warren Siegel and Edward Witten. This work was supported in part
by the Walter Burke Institute for Theoretical Physics at Caltech and by U.S. DOE
grant DE-SC0011632. The revision (v3) was completed in the summer of 2016
at the Aspen Center for Physics, which is supported by the National Science Foundation
grant PHY-1066293.

\newpage

\appendix

\section{Matrices for $SU(4)$ and $SU(2,2)$}

In order to give an economical superspace description of $AdS_5 \times S^5$ and
its $PSU(2,2|4)$ isometry, it is desirable to describe the bosonic coordinates
and the bosonic subalgebra and in an appropriate way. It is well-known that the
description of $SU(2)$ is very conveniently carried out using the three
$2 \times 2$ Pauli matrices $\s^a$. This appendix will construct $4 \times 4$
matrices, $\S^a$ and $\S^m$, that are convenient for describing $SU(4)$ and $SU(2,2)$.

In the case of $SU(4)$, we wish to define six antisymmetric $4 \times 4$ matrices
$(\S^a)^{\a\b}$ and their hermitian conjugates $(\S^{a\dagger})^{\bar\a\bar\b}$.
These matrices are invariant tensors of $SU(4)$
specifying how the six-vector representation couples to the antisymmetric Kronecker
product of two four-dimensional representations ${\bf 4} \times {\bf 4}$
and ${\bf \bar4} \times {\bf \bar4}$, respectively. An essential difference from the
case of $SU(2)$ is that the ${\bf 6}$  is not the adjoint representation of $SU(4)$.
(The latter arises in the Kronecker product ${\bf 4} \times {\bf \bar4}$.) Another
difference is that the ${\bf 4}$ representation $SU(4)$ is complex, whereas the ${\bf 2}$
representation of $SU(2)$ is pseudoreal. The invariant matrix $\eta_{\b \bar\b}$
is used to contract spinor indices in matrix products such as $\S^a \eta \S^{b\dagger}$.
However, $\eta$ is just the unit matrix $I_4$ in the case of $SU(4)$, so we
can omit it without causing confusion. In the case of $SU(2,2)$, the
matrix $\eta$ is not the unit matrix, so we will display it in this appendix,
but not in the main text.

We use the matrices $\S^a$ and $\S^{a\dagger}$ to define $4 \times 4$ matrices
\begin{equation} \label{Zdef}
Z = \vec\S \cdot \hat z \quad {\rm and} \quad Z^{\dagger} = \vec \S^{\dagger} \cdot \hat z.
\end{equation}
The six-vector $\hat z$ describes a unit five-sphere, so  $\hat z \cdot \hat z =1$.
We can encode a specific choice of the six antisymmetric matrices $(\S^a)^{\a\b}$
by introducing three complex coordinates $u = z^1 +i z^2$,
$v = z^3 + i z^4$, and $w = z^5 + iz^6$ and defining\footnote{This matrix and the one called $Y$
(below) have appeared previously in the $AdS_5 \times S^5$ literature.}
\begin{equation}\label{Zmatrix}
Z^{\a\b} = \left( \begin{array}{cccc}
0 & u & v & w  \\
-u & 0 & -\bar w & \bar v\\
-v & \bar w & 0 & -\bar u\\
- w & - \bar v & \bar u & 0 \\
\end{array} \right).
\end{equation}
It is easy to verify that this choice satisfies
\begin{equation}\label{zsquared}
Z Z^{\dagger} = Z^{\dagger} Z = I_4,
\end{equation}
which implies that $Z$ is a unitary matrix.
%Note that this matrix is antisymmetric,
%whereas the corresponding matrix in the case of $SU(2)$, $i\vec\s\cdot \vec z$, is antihermitian.

The formulas given above imply that
the $\S$ matrices satisfy the equations
\begin{equation}
(\S^a \S^{b\dagger} + \S^b \S^{a\dagger})^\a{}_{\b} = 2 \d^{ab}\d^\a_{\b}
\end{equation}
and
\begin{equation}
(\S^{a\dagger} \S^b + \S^{b\dagger} \S^{a})^{\bar\a}{}_{\bar\b} = 2 \d^{ab}\d^{\bar\a}_{\bar\b}.
\end{equation}
These imply, in particular, that
\begin{equation} \label{traceSS}
\tr(\S^a \S^{b\dagger} + \S^b \S^{a\dagger}) = 8 \d^{ab}.
\end{equation}
One can also verify that
\begin{equation}
\half \e_{\a\b\g\d} (\S^a)^{\g\d} = (\S^{a\dagger})_{\a\b},
\end{equation}
which implies that
\begin{equation} \label{Zdual}
\half \e_{\a\b\g\d} Z^{\g\d} = Z^{\dagger}_{\a\b},
\end{equation}
as expected. It is also interesting to note that
\begin{equation}
\det Z = (|u|^2 + |v|^2 + |w|^2)^2 = 1.
\end{equation}
Thus, $Z$ belongs to $SU(4)$, which means that $Z$ parametrizes $S^5$ as a subspace
of $SU(4)$. This is analogous to the equation $\s_2 \vec\s \cdot \hat x$, discussed in the
introduction, which describes $S^2$ as a subspace of $SU(2)$.
The explicit formula for $Z$, given in Eq.(\ref{Zmatrix}), is never utilized.
The purpose of presenting it is to demonstrate the existence of
matrices $\S^a$ such that $Z = \vec\S \cdot \hat z$ is an $SU(4)$ matrix.

In the introduction we interpreted the $S^2$ subspace of $SU(2)$ as an equivalence
class of $SU(2)$ matrices. Therefore it is natural to seek the
corresponding interpretation of the $S^5$ subspace of $SU(4)$. Since $Z$ is an
antisymmetric matrix, the appropriate equivalence relation
is that two elements of $SU(4)$, $g_0$ and $g_0'$, are equivalent
if and only if there exists an element $g \in SU(4)$ such that $g_0' = g^T g_0 g$.
For this choice of equivalence relation, the space of antisymmetric $SU(4)$ matrices
forms an equivalence class, and the action of an arbitrary group element $g$ on an
element $g_0$ in this class is $ g_0 \to g_0' = g^T g_0 g$. The action of the center of $SU(4)$,
which is $\IZ_4$, has a $\IZ_2$ image. If $g$ is $i$ times
the unit matrix, which is an element of the center, the map sends $g_0 \to - g_0$.
So the isometry group is really $SO(6)$, as it should be.
There are actually two $S^5$'s inside $SU(4)$, which are distinguished
by a change of sign in Eq.~(\ref{Zdual}). The map $Z \to i Z$ is a one-to-one
map relating the two spheres.

In the case of $SU(2,2)$ and $AdS_5$ we should redefine two of the six
$\S$ matrices given above by a factor of $i$ in order to incorporate the
indefinite signature of $Spin(4,2)$. Therefore we modify the $SU(4)$ formulas
accordingly and define $Y = \vec{\tilde\S} \cdot \hat y$ by
\begin{equation}
Y^{\m\n} = \left( \begin{array}{cccc}
0 & iu & v & w  \\
-iu & 0 & -\bar w & \bar v\\
-v & \bar w & 0 & -i \bar u\\
- w & - \bar v & i \bar u & 0 \\
\end{array} \right).
\end{equation}
Now, in the notation of Eq.~(1), we make the identifications
$u = y^0 +iy^5$, $v = y^1+iy^2$, and $w=y^3 +iy^4$. Since
$-y^2 = |u|^2 - |v|^2 - |w|^2 =1$ describes the Poincar\'e patch of $AdS_5$, we see
that the determinant of $Y$ is unity.
Next we take account of the indefinite signature of $SU(2,2)$ by defining
\begin{equation}
\eta^{\m\bar\n} = \eta_{\m\bar\n} = \left( \begin{array}{cc}
I_2 & 0  \\
0 & -I_2 \\
\end{array} \right) = I_{2,2}.
\end{equation}
Then, using this metric to contract spinor indices, one finds that
\begin{equation}\label{ysquared}
Y \eta Y^{\dagger} \eta =  I_4,
\end{equation}
where we use $ - y^2 = |u|^2 - |v|^2 - |w|^2 = 1$ once again.
This implies that $Y$ is an element of $SU(2,2)$. Thus, just as in the compact case,
we find that $AdS_5$ is represented as a subspace of the $SU(2,2)$ group manifold.
Eq.~(\ref{ysquared}) implies the algebra
\begin{equation}
(\S^m \eta \S^{n\dagger}\eta + \S^n \eta\S^{m\dagger}\eta)^\m{}_{\n} = - 2\eta^{mn}\d^\m_{\n},
\end{equation}
where $\eta^{mn}$ is the $SO(4,2)$ metric.

The main text takes factors of $\eta$
into account by only using ``unbarred'' indices, \ie by defining
\begin{equation}
Y^{\dagger}_{\m\n} = \eta_{\m\bar\m} Y^{\dagger \bar\m \bar\n} \eta_{\bar\n \n}.
\end{equation}
Then we can write $Y \Yd = I$, even though $Y$ is not unitary.

Other interesting quantities are the connection one-forms for $\mathfrak{su}(4)$
and $\mathfrak{su}(2,2)$. The former is given by
\begin{equation}
\O_0 = Z d Z^{\dagger} = - dZ Z^{\dagger},
\end{equation}
This matrix is antihermitian and traceless, which implies that it belongs to the $\mathfrak{su}(4)$
Lie algebra. To eliminate any possible doubt about this, we have computed the matrix
explicitly:
\begin{equation}
\O_0 = \left( \begin{array}{cccc}
u d\bar u + v d\bar v + w d\bar w & wdv - vdw & udw-wdu & vdu-udv  \\
\bar v d \bar w - \bar w d \bar v & u d \bar u + \bar v dv + \bar w dw
& u d \bar v - \bar v du & u d \bar w - \bar w du\\
\bar w d \bar u - \bar u d \bar w & v d\bar u - \bar u dv
& v d \bar v + \bar w dw + \bar u du & v d \bar w - \bar w dv\\
\bar u d\bar v - \bar v d\bar u & w d \bar u - \bar u dw
& w d \bar v - \bar v dw & wd\bar w + \bar v dv + \bar u du \\
\end{array} \right).
\end{equation}
Tracelessness is a consequence of $|u|^2 + |v|^2 + |w|^2 = 1$.
This is a flat connection, since the two-form
$d\O_0 + \O_0 \wedge \O_0$ vanishes. Similarly, the connection one-form
\begin{equation}
\tilde\O_0 = Y\eta d Y^{\dagger} \eta = - dY \eta Y^{\dagger} \eta
\end{equation}
belongs to the $\mathfrak{su}(2,2)$ Lie algebra, as it should. Moreover,
$d\tilde\O_0 + \tilde\O_0 \wedge \tilde\O_0 =0$, so it is also a flat connection.

To represent the Lie algebra of $\mathfrak{su}(4)$ we introduce the fifteen traceless antihermitian
$4 \times 4$ matrices
\begin{equation}
(\S^{ab})^{\a}{}_{\b} = \half(\S^{a} \S^{b\dagger} - \S^{b} \S^{a\dagger})^{\a}{}_{\b}.
\end{equation}
Similarly, for $\mathfrak{su}(2,2)$ we have
\begin{equation}
(\tilde\S^{mn})^{\m}{}_{\n} = \half(\tilde\S^m \eta \tilde\S^{n\dagger}\eta - \tilde\S^n \eta \tilde\S^{m\dagger}\eta)^{\m}{}_{\n}.
\end{equation}
In this notation, the representations of the $\mathfrak{su}(4)$ and
$\mathfrak{su}(2,2)$ Lie algebras are
\begin{equation}
\half [\S^{ab},\S^{cd}] =  \d^{bc}\S^{ad} + \d^{ad}\S^{bc} - \d^{ac} \S^{bd} - \d^{bd}\S^{ac}
\end{equation}
and
\begin{equation}
\half [\tilde\S^{mn},\tilde\S^{pq}] =  \eta^{np}\tilde\S^{mq} + \eta^{mq}\tilde\S^{np} - \eta^{mp} \tilde\S^{nq} - \eta^{nq}\tilde\S^{mp}.
\end{equation}

\section{Derivation of the Noether current}

The Noether procedure for constructing the conserved current associated with a global
symmetry instructs us to consider a {\em local} infinitesimal transformation, which is not a symmetry.
The variation of the action then contains the derivative of the infinitesimal parameter
times the Noether current. It then follows that conservation of the current is a consequence
of the equations of motion. We wish to apply this procedure to the action in Eq.~(\ref{action})
by considering its variation under an arbitrary {\em local} $\mathfrak{psu}(2,2|4)$
transformation $\L (\s)$ specified by the infinitesimal supermatrix
\begin{equation}
\L = \left( \begin{array}{cc}
\o &  \z \e  \\
\z \e^{\dagger} & \tilde\o \\
\end{array} \right) .
\end{equation}
The equations $\d_\L J_i = [\L,J_i]$, which are correct when $\L$ is constant,
need to be generalized to include additional terms depending on $d\L$. We
will also use the supermatrices $A_i = \G^{-1} J_i \G$, which were
introduced in Sect.~3.5.

It will prove useful to know that
\begin{equation} \label{GL}
\G^{-1} d\L  \G = \left( \begin{array}{cc}
 \chi + i\th \phi^{\dagger} -i \phi \td +i \th \tilde\chi \td
&  \z (\phi + \chi \th -\th \tilde\chi +i \th \phi^{\dagger} \th)  \\
\z (\phi^{\dagger} + \tilde\chi \td -\td \chi +i \td \phi \td)
&  \tilde\chi + i\td \phi -i \phi^{\dagger} \th +i \td \chi \th  \\
\end{array} \right) ,
\end{equation}
where
\begin{equation}
\chi = f^{-1} d\o f^{-1}, \quad \tilde\chi = {\tilde f}^{-1} d\tilde\o {\tilde f}^{-1},
\quad {\rm and} \quad \phi = f^{-1} d\e {\tilde f}^{-1} .
\end{equation}

We need to evaluate the variations of $\Str (J_1 \wedge \star J_1)$ and $\Str(J_2 \wedge J_3)$,
which are the terms that appear in the action. The key to evaluating them is to rewrite them
as $\Str (A_1 \wedge \star A_1)$ and $\Str(A_2 \wedge A_3)$ and to evaluate
\begin{equation}
\d \, \Str (A_1 \wedge \star A_1) = 2\, \Str ( \d' A_1 \wedge \star A_1)
\end{equation}
and
\begin{equation}
\d \,\Str (A_2 \wedge  A_3) = 2 \,\Str ( \d' A_2 \wedge  A_3),
\end{equation}
Because these expressions have global $\mathfrak{psu}(2,2|4)$ symmetry, we only need to keep the
terms involving $d\L$ in $\d A_i$. We have denoted these pieces by
$\d' A_i$. Calculating these, we find
\begin{equation}
\d' A_1  = -[\G^{-1} d\L \G]_{\rm even} - X [\G^{-1} d\L \G]^T_{\rm even} X^{-1},
\end{equation}
\begin{equation}
\d' A_2  = -[\G^{-1} d\L \G]_{\rm odd},
\end{equation}
\begin{equation}
\d' A_3  = i X [\G^{-1} d\L \G]^T_{\rm odd} X^{-1}.
\end{equation}
Since $A_1$ only contains diagonal blocks, which are even, so does its variation. These blocks are related to
those in $\G^{-1} d\L \G$, which is written out in Eq.~(\ref{GL}), in the indicated fashion.
Similarly, the variations of $A_2$ and $A_3$ are
given by the odd off-diagonal blocks of $\G^{-1} d\L \G$. We can add back the missing blocks
in each case, since they do not contribute to the supertraces. Therefore
\[
\d\, \Str (A_1 \wedge \star A_1) = -2 \,\Str ( \G^{-1} d\L \G \wedge \star A_1)
-2 \,\Str ( X[\G^{-1} d\L \G]^T X^{-1}\wedge \star A_1)
\]
\begin{equation}
 = - 4 \, \Str ( \G^{-1} d\L \G \wedge \star A_1)
= -4 \, \Str ( d\L \wedge \star J_1)
\end{equation}
and
\begin{equation}
\d \,\Str (A_2 \wedge  A_3) = -2 \,\Str ( \G^{-1} d\L \G  \wedge  A_3) =
-2 \,\Str ( d\L  \wedge  J_3) .
\end{equation}
Varying the combination that appears in the superstring action,
\begin{equation}
\d \,\left( \Str (A_1 \wedge \star A_1) + 2 \,\Str (A_2 \wedge  A_3) \right)
= -4 \,\Str ( d \L \wedge (\star J_1 +  J_3)).
\end{equation}
Thus, choosing the normalization, the Noether current that satisfies the conservation
equation $d \star J_N =0$ is
\begin{equation}
J_N = J_1 + \star J_3.
\end{equation}

The formulas for $\d' A_i$ imply that
\begin{equation}
\d' A_+  = -\G^{-1} d\L \G - X [\G^{-1} d\L \G]^T X^{-1}
\end{equation}
and
\begin{equation}
\d' A_-  = -\G^{-1} d\L \G - X [\G^{-1} d\L \G]^{\overline T} X^{-1}.
\end{equation}
From these it follows that
\begin{equation}
\d' J_+  = - d\L  - B_+ d\L^T B_+^{-1}
\quad {\rm and} \quad
\d' J_-  = - d\L  - B_- d\L^{\overline T} B_-^{-1},
\end{equation}
where $B_+$ and $B_-$ are defined in Eq.~(\ref{Bpm}).

\section{Kappa symmetry projection operators}

In order to figure out how kappa symmetry should work in the current context, it
is very helpful to review the flat-space limit first.
The flat-space theory was worked out using two 32-component MW spinors $\th_i$
of the same chirality \cite{Green:1983wt}.
We will summarize the results in the spinor notation of section 5.2 of \cite{Becker:2007zj},
without explaining that notation here. For an appropriate
normalization constant $k$, it was shown that the variations are
\begin{equation}
\d S_1 = k \int d^2\s \sqrt{-G} G^{\a\b} (\pa_\a \bar\th_1 \Pi_\b \r_1
+ \pa_\a \bar\th_2 \Pi_\b \r_2)
\end{equation}
where $\Pi_\a = \G_\m \Pi_\a^\m$, $G_{\a\b} = \eta_{\m\n}\Pi_\a^\m \Pi_\b^\n$, and $\r_i = \d\th_i$.
Similarly,
\begin{equation}
\d S_2 = k \int d^2\s \e^{\a\b}  (\pa_\a \bar\th_1 \Pi_\b \r_1
- \pa_\a \bar\th_2 \Pi_\b \r_2) .
\end{equation}
Despite notational differences, it should be plausible that these equations describe
the flat-space limit of the results founds in Sect.~5.1.

In this setting, the appropriate involution $\g$ turned out to be
$\g(\r) = \g\r$, where
\begin{equation} \label{gammaflat}
\g = \half \frac{\e^{\a\b}}{\sqrt{-G}} \Pi_\a^\m \Pi_\b^\n \G_{\m\n},
\end{equation}
The formula $\g^2 = I$ is equivalent to
\begin{equation}
\half\e^{\a\b} \e^{\a'\b'} \Pi^\m_\a \Pi^\n_\b \Pi^{\r}_{\a'} \Pi^{\l}_{\b'} \{\G_{\m\n},\G_{\r\l}\} = -4 G I.
\end{equation}
To prove this, note that
\begin{equation} \label{GG}
\half \{\G_{\m\n},\G_{\r\l}\} = (\eta_{\m\l}\eta_{\n\r} - \eta_{\m\r}\eta_{\n\l})I +\G_{\m\n\r\l},
\end{equation}
but the last term does not contribute, because $\a$, $\b$, $\a'$, and $\b'$ only take two values.

Another useful identity, which is proved in a similar manner, is
\begin{equation}\label{Pigamma1}
\sqrt{-G} G^{\a\b} \Pi_\b \g = \e^{\a\b} \Pi_\b .
\end{equation}
Multiplying on the right by $\g$ and using $\g^2 = I$, it is also true that
\begin{equation}\label{Pigamma2}
\sqrt{-G} G^{\a\b}\Pi_\b  = \e^{\a\b} \Pi_\b \g  .
\end{equation}
Substituting the latter identity into $\d S_1$, one obtains
\begin{equation}
\d S_1 + \d S_2 = 2k \int d^2\s \e^{\a\b}  \Big(\pa_\a\bar\th_1  \Pi_\b  \g_+ \r_1
- \pa_\a\bar\th_2 \Pi_\b \g_- \r_2 \Big)  ,
\end{equation}
where $\g_{\pm} = \half(1 \pm \g)$ are projection operators. Thus,
$\r_1 = \d\th_1 = \g_-\k$ and $ \r_2 = \d \th_2 = \g_+ \k$ are 16 local symmetries.
This means that half of the $\th$ coordinates are gauge degrees of freedom
of the string world-sheet theory.

\subsection*{The $AdS_5 \times S^5$ case}

Kappa symmetry works in a similar way for the $AdS_5 \times S^5$ background geometry.
The main challenge is to transcribe the flat-space formulas
into the matrix notation used in this manuscript.
The key equation is the defining equation of the involution $\g$. We claim
that the correct counterpart of the operator $\g$ in Eq.~(\ref{gammaflat}) is
%\begin{equation} \label{gammadef}
%\g(\r) = \half\frac{\e^{\a\b}}{\sqrt{-G}} \left( \Pi_\a \Pi_\b^{\dagger} \r
%- 2 \Pi_\a \r^{\star}\tilde\Pi_\b^{\dagger} + \r \tilde\Pi_\a \tilde\Pi_\b^{\dagger} \right).
%\end{equation}
%It can be rewritten in the more convenient form
\begin{equation} \label{gamrho}
\g(\r) = -\half\frac{\e^{\a\b}}{\sqrt{-G}} \left( \O_\a \O_\b \r
- 2 \O_\a \r' \tilde\O_\b + \r \tilde\O_\a \tilde\O_\b \right).
\end{equation}
This formula is unique up to sign ambiguities that are related to
discrete symmetries of the world-sheet theory. In particular, the sign of the second term
could be reversed. Using the definition in Eq.~(\ref{rhoprime}),
$\g(\r)$ satisfies the formula
\begin{equation}
[\g(\r)]' =  \g(\r').
\end{equation}
Therefore, if $\r$ is a MW matrix, \ie $\r = \r'$, then $\g(\r)$ is also a MW matrix.
In general we can write $\r = \r_1 + i \r_2$ and $\r' = \r_1 - i \r_2$, where $\r_1$
and $\r_2$ are MW matrices. Substituting these expressions into Eq.~(\ref{gamrho}),
it is easy to see that the general case follows from the MW case. Therefore
it is sufficient to prove that $\g \circ \g$ is the identity operator,
\ie that $\g$ is an involution, for the special case $\r' = \r$.

%Iterating the $\g$ operation generates five types
%of terms, which schematically have the structure $\O^4 \r$, $\O^3 \r' \tilde\O$,
%$\O^2 \r {\tilde\O}^2$, $\O \r' {\tilde\O}^3$, and $\r{\tilde\O}^4$. To
%understand them, recall that we need to generate a factor of the determinant of $G_{\a\b}$ to
%cancel the denominator.

The fact that $\r$ is multiplied from both the left and the right is a bit awkward.
Therefore let us recast Eq.~(\ref{gamrho}) in a form with all multiplications
acting from the left
\begin{equation}
-2 \sqrt{-G} \g(\r) = F \r,
\end{equation}
where
\begin{equation}
F = \e^{\a\b} (\O_\a \O_\b \otimes I - 2 \O_\a \otimes \tilde\O_\b^T
+ I \otimes \tilde\O_\b^T \tilde\O_\a^T) .
\end{equation}
The second factor in the tensor products acts on the second index of the matrix $\r$.
In this notation the condition that $\g$ is an involution is
\begin{equation}
F^2 = -4G (I \otimes I),
\end{equation}
which we will now verify.

In the present problem $G_{\a\b}$ is a sum of two terms, an $S^5$ part and an $AdS_5$ part,
\begin{equation}
G_{\a\b} = g_{\a\b} + {\tilde g}_{\a\b}.
\end{equation}
The crucial equations for verifying that $F^2 = -4G$ are
\begin{equation} \label{Oanti}
\{ \O_\a, \O_\b\} = - 2 g_{\a\b}\, I \quad {\rm and} \quad
\{ {\tilde\O}_\a, {\tilde\O}_\b \} = 2{\tilde g}_{\a\b} \, I .
\end{equation}
These identities are established by utilizing equations analogous
to Eq.~(\ref{GG}) for the matrices introduced in Appendix~A. They are consistent with
Eq.~(\ref{Gform}). The determinant of $G_{\a\b}$ is
the sum of three pieces: $\det g_{\a\b}$, $\det\tilde g_{\a\b}$,
and terms that are bilinear in $g_{\a\b}$ and $\tilde g_{\a\b}$. It
is now straightforward to verify that upon squaring $F$
the $\O^4 \otimes I$ terms give the $\det g$ piece, the $\O^2 \otimes {\tilde\O}^2$
terms give the mixed pieces, the $I \otimes {\tilde\O}^4$ terms give the $\det\tilde g$ piece.
Furthermore, the $\O^3 \otimes \tilde\O$ and $\O \otimes {\tilde\O}^3$ terms
vanish. Having established that $\g \circ \g = I$, we can define
orthogonal projection operators $\g_{\pm}$ by
\begin{equation}
\g_+(\r) = \half[\r + \g(\r)] \quad {\rm and} \quad \g_-(\r) =\half[\r - \g(\r)].
\end{equation}
%Though this discussion is valid for an arbitrary fermionic matrix $\r$
%that transforms as $(\bf{4, \bar4})$ under $SU(4)\times SU(2,2)$,
%the formulas will be applied to the specific matrix defined in Eq.~(\ref{rhodef}).

The $AdS_5 \times S^5$ counterpart of Eq.~(\ref{Pigamma1}) is
\begin{equation} \label{GP2}
\sqrt{-G} G^{\a\b}(\O_\b \g(\r') - \g(\r) \tilde\O_\b) =
\e^{\a\b} (\O_\b \r' - \r \tilde\O_\b).
\end{equation}
This can be proved using Eqs.~(\ref{gamrho}) and (\ref{Oanti}).
Defining a pair of one-forms,
\begin{equation} \label{pqdef}
p = \O \g(\r') - \g(\r) \tilde\O \quad {\rm and} \quad
q = \O \r' - \r \tilde\O,
\end{equation}
Eq.~(\ref{GP2}) can be recast in the more elegant form
\begin{equation} \label{pq}
p = \star q \quad {\rm or} \quad q = \star p,
\end{equation}
where the Hodge dual is defined using the induced metric $G_{\a\b}$.
This crucial identity, which is used to establish kappa symmetry in Sect.~4.4,
relates three involutions: $\star$, $\m$ (which maps $\r \to \r'$), and $\g$.

Let us now recast these results in terms of supermatrices $A_1$,
\begin{equation}
R
= \left( \begin{array}{cc}
0 & \z \r  \\
\z \r^{\dagger} & 0 \\
\end{array} \right),
\end{equation}
and
\begin{equation}
\g(R)
= \left( \begin{array}{cc}
0 & \z \g(\r)  \\
\z \g(\r)^{\dagger} & 0 \\
\end{array} \right).
\end{equation}
Corresponding to $\r=\r_1 +i\r_2$ and $\r'=\r_1-i\r_2$,
we can write $R=R_1 + i R_2$ and $R' = R_1 -i R_2$, where $R_1$ and
$R_2$ satisfy $R_i = i X R_i^T X^{-1}$.
For MW supermatrices, such as $R_1$ and $R_2$, Eqs.~(\ref{pqdef}) and (\ref{pq}) combine to give
\begin{equation} \label{keyid}
[ R_i, A_1] = [ \g(R_i), \star A_1] \quad i=1,2.
\end{equation}
Together with the fact that $\g_+$ and $\g_-$ are orthogonal projection operators,
Eq.~(\ref{keyid}) is the key formula that is utilized in the proof of kappa symmetry
in Sect.~4.4.

\end{document}